\documentclass[journal,final]{IEEEtran}
\IEEEoverridecommandlockouts
\usepackage{cite}
\usepackage{amsmath,amssymb,amsfonts,amsthm}
\usepackage{algorithmic}
\usepackage{algorithm}
\usepackage{graphicx}
\usepackage{textcomp}
\usepackage{xcolor}
\usepackage{url}
\usepackage{lipsum}
\usepackage{multirow}
\usepackage[normalem]{ulem}
\usepackage{bbding}
\usepackage{pifont}
\usepackage{wasysym}
\usepackage{bbm}
\usepackage{dsfont}
\usepackage{mathtools, nccmath}
\usepackage{makecell}
\usepackage{aligned-overset}
\usepackage{stfloats}

\DeclareMathAlphabet{\mathbcal}{OMS}{cmsy}{b}{n}
\def\BibTeX{{\rm B\kern-.05em{\sc i\kern-.025em b}\kern-.08em
		T\kern-.1667em\lower.7ex\hbox{E}\kern-.125emX}}
\newtheorem{remark}{Remark}

\newtheorem{lemma}{Lemma}
\newtheorem{corollary}{Corollary}
\newtheorem{definition}{Definition}


\begin{document}
	\title{Near-Field Communication with Massive Movable Antennas: A Functional Perspective\\
	}
	\author{Shicong~Liu,~\IEEEmembership{Graduate Student Member,~IEEE}, Xianghao~Yu,~\IEEEmembership{Senior Member,~IEEE}, Jie~Xu,~\IEEEmembership{Fellow,~IEEE}, and Rui Zhang,~\IEEEmembership{Fellow,~IEEE}
		\thanks{
			
			Shicong Liu and Xianghao Yu are with the Department of Electrical Engineering, City University of Hong Kong, Hong Kong (email: sc.liu@my.cityu.edu.hk, alex.yu@cityu.edu.hk).
			
			Jie Xu is with the School of Science and Engineering and Future Network of Intelligence Institute (FNii), The Chinese University of Hong Kong (Shenzhen), Shenzhen, China (email: xujie@cuhk.edu.cn).
			
			Rui Zhang is with the Department of Electrical and Computer Engineering, National University of Singapore, Singapore 117583 (email: elezhang@nus.edu.sg).
			
		}
	}
	
	\maketitle
	
	\begin{abstract}
		The advent of massive multiple-input multiple-output (MIMO) technology has provided new opportunities for capacity improvement via strategic antenna deployment, especially when the near-field effect is pronounced due to antenna proliferation. 
		In this paper, we investigate the optimal antenna placement for maximizing the achievable rate of a point-to-point near-field channel, where the transmitter is deployed with massive movable antennas. First, we propose a novel design framework to explore the relationship between antenna positions and achievable data rate. 
		By introducing the continuous antenna position function (APF) and antenna density function (ADF), we reformulate the antenna position design problem from the discrete to the continuous domain, which maximizes 
		the achievable rate functional with respect to ADF. 
		Leveraging functional analysis and variational methods, we derive the optimal ADF condition and propose a gradient-based algorithm for numerical solutions under general channel conditions. 
		Furthermore, for the near-field line-of-sight (LoS) scenario, we present a closed-form solution for the optimal ADF, revealing the critical role of edge antenna density in enhancing the achievable rate. Finally, we propose a flexible antenna array-based deployment method that ensures practical implementation while mitigating mutual coupling issues. Simulation results demonstrate the effectiveness of the proposed framework, with uniform circular arrays emerging as a promising geometry for balancing performance and deployment feasibility in near-field communications.
	\end{abstract}
	
	
	\section{Introduction}
	\bstctlcite{IEEEexample:BSTcontrol}

	The evolution of beyond fifth-generation (B5G) and forthcoming sixth-generation (6G) wireless networks has ushered tremendous demands for higher data rates, ultra reliability, and massive device connectivity~\cite{10379539}. To meet these stringent requirements, unlocking and efficiently utilizing spatial degrees of freedom (DoFs) have become a key focus in improving the performance of wireless communication systems. 
	Among various enabling technologies, advanced antenna technologies, especially massive multiple-input multiple-output (MIMO), have demonstrated great potential in boosting spatial multiplexing efficiency. Building on massive MIMO, large-scale reconfigurable antenna technologies, such as reconfigurable intelligent surfaces (RIS)~\cite{10555049}, reconfigurable holographic surfaces (RHS)~\cite{9696209}, and dynamic metasurface antennas (DMA)~\cite{9324910}, offer energy-efficient and cost-effective ways to 
	further expand the spatial multiplexing capabilities~\cite{7397861,bjornson24spatialmul}.

	In conventional reconfigurable antenna arrays that are passive or active, radiation elements are typically deployed at fixed positions, with uniform spacing being the most widely-adopted configuration~\cite{10555049,9696209,9324910}. While such arrays are well-suited for specific scenarios, such as uniform rich scattering environments~\cite{10.5555/1111206} under far-field conditions, their performance is limited in more complex scenarios. For instance, in millimeter-wave communications, scatterers tend to exhibit clustered distributions rather than a uniform spatial spread~\cite{7397861,7546944}. Furthermore, as the aperture of massive reconfigurable antenna arrays increases, the near-field effect becomes more significant~\cite{10220205}, causing the transition of incident electromagnetic (EM) wavefronts from planar to spherical. In this case, the conventional fixed-position arrangement of array elements is no longer sufficient to unleash full spatial DoFs under the spherical wavefront in near-field wireless communications~\cite{10446604,10909572}. %

	With recent advancements in mechanical innovation and antenna design, position-reconfigurable antenna (PRA) technologies, such as movable antenna (MA)~\cite{10286328} and fluid antenna systems~\cite{9264694}, have gained significant attention from both academia and industry~\cite{10278220,10416363}. 
	Specifically, the MA architecture achieves position reconfiguration by mechanically moving antenna elements~\cite{10286328} or sub-arrays~\cite{10851455}, while the fluid antenna architecture dynamically selects the strongest ports using liquid metals~\cite{9264694}. 
	These systems allow for more favorable propagation environment and precise manipulation of radiation patterns, thereby improving overall communication performance. 
	As the massive MIMO technology becomes mainstream and widely deployed, there is an urgent need for optimal design strategies tailored to reconfigurable massive movable antenna systems in the near-field region.

	\subsection{Related Works}
	
	PRAs have validated their considerable performance gains compared to conventional fixed-position antennas for 
	both communication and sensing in modern wireless systems~\cite{10328751,10504625,10354003,10851455,10416363,10684758,10476966,10643473,10243545}. 
	The performance advancements span a wide range of metrics, including achievable rates~\cite{10328751,10504625,10354003,10851455}, physical layer security~\cite{10416363,10684758}, sensing Cram\'er-Rao bound (CRB)~\cite{10476966,10643473}, etc. 
	Specifically, a method for optimizing the upper bound of achievable rates was proposed in~\cite{10328751} for fluid antenna-assisted systems, where antenna positions and precoding vectors are determined by optimizing the transmit covariance matrix. Simulation results therein demonstrated that the achievable rate increases with the moving region and quickly reaches saturation, indicating that a limited aperture is sufficient to achieve the maximum achievable rate. In addition, the weighted sum-rate maximization problem in MA-enhanced multiuser MIMO systems was investigated in~\cite{10504625}, where the problem was reformulated into a weighted minimum mean square error (WMMSE) form for tractable optimization. 
	A block coordinate descent (BCD)-based method with a constrained antenna movement strategy was further proposed to reduce the computational complexity by $30\%$ without significant performance degradation. 
	Furthermore, the authors of~\cite{10851455} addressed the sum-rate maximization problem with the sub-connected architecture, where the maximal sum-rate performance is achieved by jointly optimizing the digital beamformer, analog beamformer, and movable subarrays' positions. Fractional programming and alternating optimization frameworks were introduced to tackle the non-convexity of the optimization problem, whereas simulation results illustrated the superiority of movable sub-connected arrays compared to conventional fixed-position antenna arrays.
	
	Apart from the advancements in communication performance, recent research has also explored the potential of PRA arrays to enhance physical layer security and sensing capabilities. 
	The secrecy rate maximization problem was studied in the presence of multiple single-antenna and colluding eavesdroppers~\cite{10416363}, where the secrecy rate was maximized by jointly designing the beamformer and positions of all antennas at the transmitter. Projected gradient ascent and alternating optimization were employed to address the non-convexity of the problem and obtain high-quality sub-optimal solutions. Besides, a design  framework was proposed in~\cite{10684758} to maximize the secrecy rate by reformulating the problem into surrogate tractable forms, leveraging the majorization-minimization algorithm to iteratively update the transmit beamforming matrix and antenna positions. In particular, PRA arrays demonstrated a remarkable capability to suppress sidelobes in desired directions and achieved superior secrecy rate performance compared to conventional fixed-position antennas in both works. 
	Furthermore, since CRB is inherently dependent on antenna positions, optimizing the worst-case CRB~\cite{10476966} or directly minimizing CRB under practical constraints~\cite{10643473} has been investigated to improve the accuracy of direction of arrival (DoA) estimation in wireless sensing scenarios. 
	Recently, PRA has also been extended to the general six-dimensional MA (6DMA) architecture enabling both three-dimensional (3D) antenna position and 3D antenna rotation adjustments~\cite{10752873,10848372,10883029,10945745}. 
	
	\subsection{Motivations}
	
	Despite the various benefits achieved through optimizing antenna positions, the prior works still face several significant limitations. First, existing works heavily rely on non-convex optimization, which leads to the fact that the developed design algorithms involve iterative procedures with nested loops. Consequently, their computational complexity can become extremely high, as it typically grows polynomially with the number of antenna elements. Therefore, in practice these algorithms are only applicable to wireless systems with a limited number of antennas~\cite{10328751,10504625,10354003,10851455,10416363,10684758,10476966,10643473,9264694,10092780,10278220}, which deviates from the massive MIMO evolution anticipated in B5G/6G wireless communication systems. 
	
	Besides, prior works have primarily treated the antenna placement as mathematical optimization problems~\cite{10354003,10278220,10416363,10684758,10476966,10643473,9264694,10092780,10851455,10504625}, while they lack analytical approaches which can provide new and deep insights into optimal system design. 
	For instance, it is difficult to determine the optimal system parameters (e.g., the maximum moving region and the optimal positions) by following a consistent approach in different application scenarios and system setups, which significantly limits the practical application of existing results.

	Lastly, the moving area introduced by PRA arrays results in a larger array aperture, which drastically expands the near-field region~\cite{10220205,10845870,9693928}. However, most existing works on PRA did not dedicatedly take near-field effects into consideration. Moreover, as the spherical wave model under near-field conditions introduces a more intricate EM environment, it is necessary to further employ PRA arrays to fully exploit the spatial DoFs in the near-field region. 
	In summary, a low-complexity yet generally applicable design approach for PRA-based near-field massive MIMO systems is still missing in the literature, to the authors' best knowledge.

	\subsection{Contributions}

	In this paper, we investigate the optimal antenna placement in a point-to-point near-field massive MIMO system, where the transmitter is equipped with massive movable antennas. 
	To characterize the achievable rate performance, we assume that the transmitter perfectly knows the channel state information (CSI). 
	A design framework is accordingly devised to reveal the relationship between the antenna positions and the achievable rate of near-field channels. 
	Our contributions are summarized as follows:
	\begin{itemize}
		\item We propose a novel framework to formulate and address the antenna placement problem in the continuous domain. Specifically, we first introduce the continuous antenna position function (APF) and antenna density function (ADF), and reveal their inherent relationship, based on which the antenna positions are determined with negligible computational complexity. 
		Then, with the proposed framework, we reformulate the achievable rate maximization problem from the discrete domain to the continuous domain, which simplifies the problem and enables the derivation of the asymptotically optimal ADF that maximizes the rate functional.
		\item By leveraging the functional analysis, we employ the variational method to derive the optimal condition of ADF that maximizes the achievable rate. Furthermore, we propose a variational gradient-based method to numerically solve the achievable rate maximization problem, which is applicable to arbitrary channel conditions.
		\item For the near-field communication scenario, we further derive an asymptotic \textit{closed-form} solution for the optimal ADF and reveal that the edge density of antennas plays a crucial role in maximizing the achievable rate of near-field line-of-sight (LoS) channels. Particularly, a greater number of pole-type singularities and higher orders of the singularity in ADF can effectively improve the achievable rate. 
		\item We finally propose 
		a flexible antenna array-based method for deploying antenna elements, which ensures that the projected antenna density closely approximates the derived ADF function while avoiding mutual coupling issues caused by excessively small antenna spacing. Additionally, uniform circular arrays are demonstrated to be a favorable antenna geometry in near-field communications, as it strikes a delicate balance between achievable performance and practical deployment.
	\end{itemize}

\par {\it Notations}: We use normal-face letters to denote scalars and lowercase (uppercase) boldface letters to denote column vectors (matrices). The $k$-th row vector and the $m$-th column vector of matrix ${\bf H}\in\mathbb{C}^{K\times M}$ are denoted as ${\bf H}[{k,:}]$ and ${\bf H}[{:,m}]$, respectively, and the $n$-th element in the vector $\bf h$ is denoted by ${\bf h}[n]$. $\{{\bf H}_n\}_{n=1}^N$ denotes a matrix set with the cardinality of $N$. 
The superscripts $(\cdot)^{T}$, $(\cdot)^{\rm *}$, and $(\cdot)^{H}$ represent the transpose, conjugate, and conjugate transpose operators, respectively. $\det(\cdot)$, ${\rm Tr}(\cdot)$ and $\Lambda^{(i)}\left(\cdot \right)$ denote the determinant, trace, and the $i$-th eigenvalue of a matrix, respectively. 
$\mathbb{C}$, $\mathbb{R}$, and $\mathbb{Z}$ denote the set of complex numbers, real numbers and integers, respectively. $\Re(\cdot)$ and $\Im(\cdot)$ denote the real and imaginary parts of a complex number, respectively, and the imaginary unit is represented as $\jmath$ such that $\jmath^2=-1$. We use $\mathcal{O}(\cdot)$ to represent the big-$O$ notation.
\section{System Model and Problem Formulation}

In this section, we present the considered system model for antenna position design, including the reconfigurable array structure and the channel model, and then formulate the achievable rate maximization problem.
\subsection{Position-Reconfigurable (Movable) MIMO Antennas}
\begin{figure}
	\centering
	\includegraphics[width=0.45\textwidth]{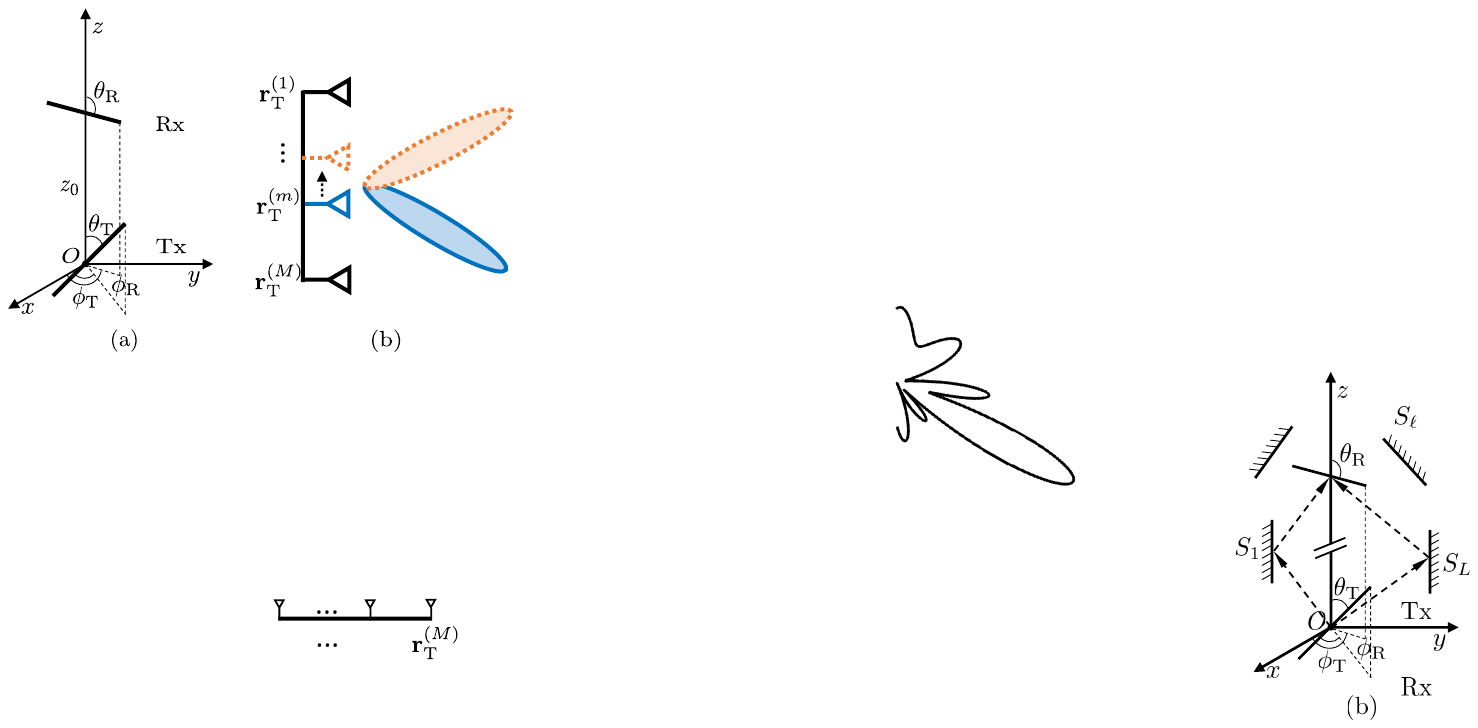}
	\caption{(a) The considered near-field communication scenario under the Cartesian coordinate system, with the $z$-axis aligned along the line connecting the centroids of the transmit and receive antenna arrays, and (b) the schematic diagram of massive movable antennas, with the $1$-st and $M$-th antenna fixed at the edges.}
	\label{fig:sysmodel}
	
\end{figure}
We consider the downlink transmission in a point-to-point massive MIMO system with $M$ movable antennas at the base station (BS) and $N$ fixed-position antennas\footnote{To keep the presentation neat and promote practicality, in this paper we consider movable antennas at the BS, as movable antenna elements are less implementable on compact UE arrays. However, due to the spatial duality of the channel, the proposed methodology in this paper can also be extended to UEs with movable antennas.} at the user equipment (UE). 
The Cartesian coordinate system is established such that the centroids of the transmit and receive antenna arrays are located at the origin and $(0,0,z_0)$, respectively. 
Antenna elements are assumed to move freely on the array. Specifically, for linear arrays, as shown in Fig.~\ref{fig:sysmodel}, the coordinate $\mathbf{r}_\mathrm{T}^{(m)}=(x_{\rm T}^{(m)},y_{\rm T}^{(m)},z_{\rm T}^{(m)} )$ of the $m$-th antenna at the BS side can be defined by parametric equations as
\begin{equation}
	\begin{cases}
		x_{\rm T}^{(m)} \!&= \frac{A_{\rm T}}{2} f(m)  \sin\theta_{\rm T}\cos\phi_{\rm T}  \\
		y_{\rm T}^{(m)} \!&= \frac{A_{\rm T}}{2} f(m) \sin\theta_{\rm T}\sin\phi_{\rm T},\quad \forall m\in\mathcal{M},  \\
		z_{\rm T}^{(m)} \!&= \frac{A_{\rm T}}{2} f(m) \cos\theta_{\rm T}
	\end{cases}
	\label{eq:ant_loc}
\end{equation}
where $\mathcal{M}=\{m\in \mathbb{Z} \mid 1\leq m\leq M\}$, and $A_{\rm T} = (M-1)d$ denotes the standard aperture of the BS array with $d$ denoting the unit antenna spacing for the array. 
In addition, $\phi_{\rm T}$ and $\theta_{\rm T}$ are the azimuth and elevation angles of the BS array, respectively. 
$f(m)$ denotes the antenna position function (APF) for the $m$-th antenna element. For example, the APF for the conventional uniform linear array (ULA) is given by
\begin{equation}
	f^{\rm uni}(m) = \frac{2}{M-1}\left( m - \frac{M+1}{2} \right).%
	\label{eq:uni}
\end{equation}
In this paper, the APF $p=f(m)$ of the antenna elements increases in ascending order with the antenna index $m$, i.e., the APF $f(m)$ is a monotonically increasing function bounded by $-1\leq f(m)\leq 1$, which ensures that the overall aperture\footnote{The APF model can adapt to different array apertures by modifying the standard antenna spacing $d$.} is bounded by $\Vert {\bf r}_{\rm T}^{(M)}- {\bf r}_{\rm T}^{(1)} \Vert = A_{\rm T}$.

\subsection{Problem Formulation}

In this paper, to characterize the fundamental limit of MA-assisted near-field communications, we assume that perfect CSI is available. In practice, the BS can first adopt a fixed array configuration, e.g., ULA, to facilitate a standard channel estimation~\cite{10497534,10659325,10643473}, in which key channel parameters, e.g., locations of scatterers and UE, channel gains, are accurately estimated. With the perfect CSI at hand, we aim to maximize the achievable rate by further deriving the optimal positions of MA elements. 

For an arbitrary channel matrix ${\bf H}_f\in\mathbb{C}^{N\times M}$ modeled by
\begin{equation}
	{\bf H}_f[n,m] = h\left({\bf r}_{\rm R}^{(n)},{\bf r}_{\rm T}^{(m)}\right),
	\label{eq:arb_chan}
\end{equation}
where $h({\bf r}_{\rm R}^{(n)},{\bf r}_{\rm T}^{(m)})$ denotes the channel spatial response function from the $m$-th transmitting antenna at ${\bf r}_{\rm T}^{(m)}$ to the $n$-th receiving antenna at ${\bf r}_{\rm R}^{(n)}$, and $f$ is the APF defined in~\eqref{eq:ant_loc} that controls the position of antenna elements ${\bf r}_{\rm T}^{(m)}$ at the BS array. Hence, the downlink channel matrix is a functional of APF, and the achievable rate is given by
\begin{equation}
	\begin{aligned}
		C_f=\log_2\det \left( {\bf I} + \frac{1}{\sigma_{\rm n}^2} {\bf H}_f^{} {\bf Q} {\bf H}_f^H \right)%
		\label{eq:cap}
	\end{aligned}
\end{equation}
where $\sigma_{\rm n}^2$ is the power of the additive white Gaussian noise (AWGN) signal and ${\bf Q}$ denotes the transmit covariance matrix. Since this work mainly focuses on investigating the performance gain brought by MA, we assume an isotropic transmission with ${\bf Q}= \frac{P_{\rm T}}{M}{\bf I}$ in this paper, where $P_{\rm T}$ denotes the power of the transmitted signal. One can always apply the singular value decomposition (SVD) and water-filling power allocation to obtain the optimal covariance matrix $\mathbf{Q}$ once the antenna positions are determined. Therefore, the achievable rate is further given by
\begin{equation}
	C_f=\sum_{i}^{N} \log_2\left( 1 + \rho \Lambda^{(i)}\left( {\bf K}_f\right) \right),
\end{equation}
where $\rho = P_{\rm T}/\sigma_{\rm n}^2$ denotes the signal-to-noise ratio (SNR), and the Gram matrix ${\bf K}_f$ of channel ${\bf H}_f$ is defined as
\begin{equation}
	{\bf K}_f=\frac{1}{M}{\bf H}_f^{}  {\bf H}_f^H. %
	\label{eq:gram}
\end{equation}
In addition, $\Lambda^{(i)}\left({\bf K}_f\right)$ is the $i$-th eigenvalue of the Gram matrix ${\bf K}_f$. Our target is to find the optimal APF $f(m)$ such that by reorganizing the intrinsic structure of the Gram matrix ${\bf K}_f$, more orthogonal transmission modes can be excited to achieve a higher value of rate functional. In other words, the overall problem can be formulated as
\begin{equation}
	\mathcal{P}_1:~
	\begin{aligned}
		&\underset{\{f(m)\}_{m\in\mathcal{M}}}{\max}\!\!\!\! && C_f\\
		&~~~~~\mathrm{s.t.}&&
		f(m+1) > f(m),~m\in\mathcal{M},\\
		& \!\!&&f(M)=-f(1) = 1,
	\end{aligned}
	\label{eq:P1}
\end{equation}
where $f(m+1)>f(m)$ indicates the monotonically increasing property of the APF $f(m)$, while $f(M)=-f(1) = 1$ specifies the range of $f(m)$ over $m\in\mathcal{M}$.
\section{Optimal Antenna Position Design}

The problem in~\eqref{eq:P1} is formulated based on discrete antenna indices in $\mathcal{M}$, which can lead to heavy computational complexity that increases at least polynomially with the number of antennas $M$ and provides limited insight into the antenna position design~\cite{10328751,10504625,10354003,10851455,10416363,10684758,10476966,10643473,9264694,10092780,10278220}. 
To tackle these issues, in this section we introduce the antenna density function to reformulate the problem in a continuous form and then derive the corresponding optimality condition given arbitrary channel responses.

\subsection{Antenna Density Function}
\label{sec:adf}

\begin{figure}
	\centering
	\includegraphics[width=0.45\textwidth]{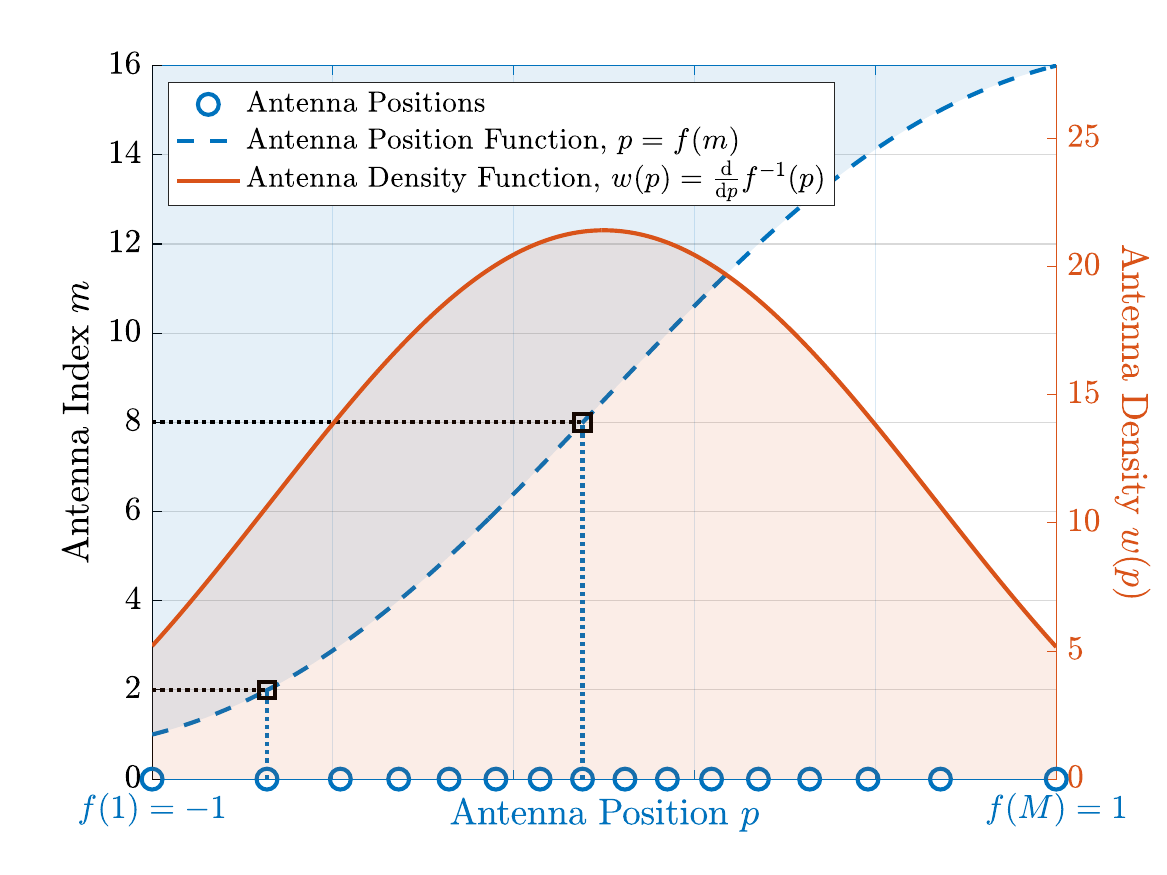}
	\caption{The APF $p = {f}(m)$ (in blue) and ADF $w(p)$ (in orange) for an $M=16$-element movable antenna array. 
	}
	\label{fig:measure}
\end{figure}

We first extend the domain of APF $f(m)$ from $m\in \mathcal{M}$ to $m\in \mathcal{M}_{\mathrm{c}} = [1,M]$ to facilitate a more tractable analysis. {By extending the domain of function $f(m)$, we assume that the antenna elements are continuously distributed over $p\in[-1,1]$, i.e., the aperture area.} 
Therefore, for an arbitrary APF on a given antenna array\footnote{The proposed scheme in this paper can also be extended to antenna arrays with arbitrary geometric shapes by modifying the parametric position equation~\eqref{eq:ant_loc}, as long as the APF $f(\cdot)$ is a bijective function.}, we introduce the concept of \textit{antenna density function} (ADF), i.e., the number of antenna elements within unit length, as
\begin{equation}
	\begin{aligned}
		w(p) = \lim_{\Delta p\to 0} \frac{f^{-1}\left(p + \Delta p\right)-f^{-1}\left(p \right)}{\Delta p} = \frac{\rm d}{{\rm d} p} {f}^{-1}(p),
	\end{aligned}
	\label{eq:adfdef}
\end{equation}
where $m={f}^{-1}(p)$ with $p\in\left[-1,1\right]$ indicates the continuous antenna index given the normalized antenna position $p$ on the array, and the inverse of $f(m)$ exists since the APF is monotonically increasing and bijective. A sample visualized relationship between APF $f(m)$ and its corresponding ADF $w(p)$ is shown in Fig.~\ref{fig:measure}. In particular, the antenna elements are denser around the center with larger values of ADF, while sparsely deployed at the edges. Compared to the APF, ADF can provide direct insights into the distribution of antenna elements on the antenna array.

It should be noted in~\eqref{eq:ant_loc} that APF $f(m)$ introduces non-linearity between the spatial coordinates ${\bf r}_{\rm T}^{(m)}$ and antenna index $m$, which makes theoretical analysis and optimization over the discrete  antenna positions less tractable, especially when $f(m)$ takes a complicated form. To address this issue, in the following we propose to reformulate problem $\mathcal{P}_1$ by adopting ADF $w(p)$ as the optimization variable. To ensure the consistency of constraints, according to~\eqref{eq:adfdef}, the monotonicity of $f(m)$ guarantees the strict increasing of its inverse $m = f^{-1}(p)$, which yields $w(p)>0$. Besides, as the ADF is tightly supported within the normalized aperture area $p\in[-1,1]$, we should have the quantity constraint given by
\begin{equation}
	\int_{-1}^1 w(p) \mathrm{d} p=\left.f^{-1}(p)\right|_{-1} ^1=M-1.
	\label{eq:pwr_constraint}
\end{equation}

According to~\eqref{eq:arb_chan} and~\eqref{eq:gram}, the $(n,n^\prime)$-th element of the Gram matrix ${\bf K}_f$ is expressed by
\begin{equation}
	{\bf K}_f[n,n^\prime]= \frac{1}{M} \sum_{m=1}^{M} h\left({\bf r}_{\rm R}^{(n)},{\bf r}_{\rm T}^{(m)}\right)h^*\!\left({\bf r}_{\rm R}^{(n^\prime)},{\bf r}_{\rm T}^{(m)}\right).
\end{equation}
Recall that the coordinates ${\bf r}_{\rm T}^{(m)}$ are critically determined by the continuous APF $p=f(m)$ in~\eqref{eq:ant_loc}. Since the domain of APF $f(m)$ has been extended from the discrete $\mathcal{M}$ to continuous $\mathcal{M}_{\rm c}$, the $(n,n^\prime)$-th element of the Gram matrix is then rewritten by taking the limit form of summation as an integral~\cite{abramowitz1948handbook} as
\begin{align}
	\tilde{\bf K}_f[n,n^\prime] &= \int_{1}^{M} h\left({\bf r}_{\rm R}^{(n)},{\bf r}_{\rm T}^{(m)}\right)h^*\!\left({\bf r}_{\rm R}^{(n^\prime)},{\bf r}_{\rm T}^{(m)}\right)\,{\rm d}m.\label{eq:gram2}
\end{align}
We then employ an integral by substitution with $m=f^{-1}(p)$ and further obtain
\begin{align}
	\tilde{\bf K}_f[n,n^\prime]
	={}&\int_{-1}^{1} h\left({\bf r}_{\rm R}^{(n)},\tilde{\bf r}_{{\rm T}}^{p}\right) h^*\left({\bf r}_{\rm R}^{(n^\prime)},\tilde{\bf r}_{{\rm T}}^{(p)}\right)\frac{\rm d}{{\rm d} p} {f}^{-1}(p)\,{\rm d}p\notag\\
	={}&\int_{-1}^{1} w(p)h\left({\bf r}_{\rm R}^{(n)},\tilde{\bf r}_{{\rm T}}^{(p)} \right) h^*\left({\bf r}_{\rm R}^{(n^\prime)},\tilde{\bf r}_{{\rm T}}^{(p)} \right)\,{\rm d}p\notag\\
	\triangleq{}&\tilde{\bf K}_w[n,n^\prime],\label{eq:kernel_p}
\end{align}
where $\tilde{\bf r}_{{\rm T}}^{(p)} = (\tilde{x}_{\rm T}^{(p)},\tilde{y}_{\rm T}^{(p)},\tilde{z}_{\rm T}^{(p)})$ is a coordinate parameterized linearly by position $p\in[-1,1]$ as
\begin{equation}
	\begin{cases}
		\tilde{x}_{\rm T}^{(p)} \!&= \frac{A_{\rm T}}{2} p \sin\theta_{\rm T}\cos\phi_{\rm T}  \\
		\tilde{y}_{\rm T}^{(p)} \!&= \frac{A_{\rm T}}{2} p \sin\theta_{\rm T}\sin\phi_{\rm T},\quad \forall p\in[-1,1].  \\
		\tilde{z}_{\rm T}^{(p)} \!&= \frac{A_{\rm T}}{2} p \cos\theta_{\rm T}
	\end{cases}
	\label{eq:ant_loc2}
\end{equation}
Hence, the problem in~\eqref{eq:P1} can be reformulated as
\begin{equation}
	\mathcal{P}_2:\quad
	\begin{aligned}
		&\underset{w(p)}{\max} && C_w\\
		&\mathrm{s.t.}&&w(p)> 0,\\
		& &&\int_{-1}^{1} w(p)\,{\rm d}p = M-1,
	\end{aligned}
	\label{eq:P2}
\end{equation}
where the rate functional is given by
\begin{equation}
	C_{w} = \log_2 \det \left( \mathbf{I} + \rho \tilde{\bf K}_w \right).
	\label{eq:cap_func}
\end{equation}

\begin{figure*}[b]
	\centering
	\rule{\textwidth}{0.5pt}
	\begin{equation}
		\begin{aligned}
			\delta C_w&=\frac{\rho}{\log 2}{\rm Tr}\left( {\bf G}_w \int_{-1}^{1}  v(p) {h\left({\bf r}_{\rm R}^{(n)},\tilde{\bf r}_{{\rm T}}^{(p)}\right) h^*\left({\bf r}_{\rm R}^{(n^\prime)},\tilde{\bf r}_{{\rm T}}^{(p)}\right)}{\rm d}p \right)\\
			&=\frac{\rho}{\log 2} \int_{-1}^{1}v(p)\left( \sum_{n=1}^N\sum_{n^\prime=1}^N {\bf G}_w[n,n^\prime] 
			{h\left({\bf r}_{\rm R}^{(n)},\tilde{\bf r}_{{\rm T}}^{(p)}\right) h^*\left({\bf r}_{\rm R}^{(n^\prime)},\tilde{\bf r}_{{\rm T}}^{(p)}\right)} \right){\rm d}p
		\end{aligned}
		\label{eq:deltaCw2}\tag{22}
	\end{equation}
\end{figure*}

\begin{remark}
	The advantages of introducing the ADF $w(p)$ as the optimization variable, instead of the APF $f(m)$, are twofold. First, the antenna coordinates ${\bf r}_{\rm T}^{(m)}$ change from a non-linear form~\eqref{eq:ant_loc} with respect to variable $m$ to a linear one~\eqref{eq:ant_loc2} with $p$. While this transformation seems elementary, it effectively simplifies the theoretical analysis and enables more direct insights that are typically unavailable under discrete settings with APF, as will be presented in Section~\ref{sec:closedform}. Second, the reformulation in~\eqref{eq:P2} is developed based on the transformation of the optimization domain from the discrete set $m\in {\mathcal M}$ to a continuous interval $p\in[-1,1]$, thereby allowing the design of 
	more efficient algorithms that inherit system insights. 
	Furthermore, once the ADF is properly determined, a practical antenna deployment strategy can be readily obtained given the relationship between the ADF and APF, as shall be illustrated in Section~\ref{sec:discrete}.
\end{remark}
\noindent However, unlike existing works that directly treat the position $\mathbf{r}_\mathrm{T}^{(m)}$ as the design parameter~\cite{10328751,10504625,10354003,10851455,10416363,10684758,10476966,10643473,9264694,10092780,10278220}, the introduction of the ADF converts the achievable rate into a functional $C_w$. This transformation calls for new techniques in solving problem $\mathcal{P}_2$, which is the main task to be addressed in the remainder of this paper.

\subsection{Variational Gradient Ascent for Optimal ADF}

Before we proceed to investigate the optimal ADF for near-field communications, in the following subsection we first present a general variational gradient ascent method to maximize the rate functional with respect to ADF under arbitrary channel responses.

To address problem $\mathcal{P}_2$ in~\eqref{eq:P2}, we first introduce the Lagrangian form that incorporates the constraints in~\eqref{eq:P2} by
\begin{align}
	&\mathcal{L}_w(\varpi,\varsigma)\\ 
	={}&C_w+\varpi \left(M-1-\int_{-1}^{1}  w(p){\rm d}p  \right)-\int_{-1}^{1}\varsigma(p) w(p){\rm d}p,\notag
	\label{eq:lag}
\end{align}
where $\varpi$ is the Lagrange multiplier, and $\varsigma(p)$ is the multiplier function associated with the non-negativity constraint. As the complementary slackness implies $\varsigma(p)=0$ for all $w(p)>0$, the optimal condition for problem $\mathcal{P}_2$ is then given by
\begin{equation}
	\frac{\delta \mathcal{L}_w(\varpi,\varsigma)}{\delta w(p)} = \frac{\delta {C}_w}{\delta w(p)} - \varpi = 0,
\end{equation}
where ${\delta {C}_w}/{\delta w(p)}$ denotes the functional derivative of $C_w$ with respect to $w$. 
More specifically, consider a small functional perturbation $\delta w(p) = \epsilon v(p)$ on $w(p)$ as
\begin{equation}
	w(p)\rightarrow w(p) + \epsilon v(p),
\end{equation} 
where $\epsilon$ is an infinitesimal parameter, and $v(p)$ is an arbitrary perturbation function that satisfies
\begin{equation}
	\int_{-1}^{1} v(p)\,{\rm d}p = 0
\end{equation}
to preserve the normalization constraint in~\eqref{eq:P2}. Our target is to find the condition that for an arbitrary perturbation $v(p)$, the variation of ${C}_w$ is $\delta {C}_w = 0$. 
The first-order variation of ${C}_w$ with respect to $w$ is defined as
\begin{align}
		\!\!\!\!\delta {C}_w &= \left.\frac{{\rm d}}{{\rm d} \epsilon} C_{w+\epsilon v} \right|_{\epsilon = 0}\!\!=\left.\frac{{\rm d}}{{\rm d} \epsilon} \log_2 \det \left( \mathbf{I} + \rho \tilde{\mathbf K}_{w+\epsilon v} \right) \right|_{\epsilon = 0}\notag\\
		&=\left.\frac{\rho}{\log 2}{\rm Tr}\left( \left( \mathbf{I} + \rho \tilde{\mathbf K}_w \right)^{-1} \frac{{\rm d}}{{\rm d} \epsilon} \tilde{\mathbf K}_{w+\epsilon v} \right)  \right|_{\epsilon = 0}.
	\label{eq:deltaCw}
\end{align}
According to~\eqref{eq:kernel_p}, for the Gram matrix $\tilde{\mathbf K}_{w+\epsilon v}$, we further have
\begin{equation}
	\begin{aligned}
		&\frac{{\rm d}}{{\rm d} \epsilon} \tilde{\mathbf K}_{w+\epsilon v}[n,n^\prime] \\
		={}& \int_{-1}^{1}  v(p) {h\left({\bf r}_{\rm R}^{(n)},\tilde{\bf r}_{{\rm T}}^{(p)}\right) h^*\left({\bf r}_{\rm R}^{(n^\prime)},\tilde{\bf r}_{{\rm T}}^{(p)}\right)}\,{\rm d}p.
	\end{aligned}
	\label{eq:deltaKw}
\end{equation}
Substituting~\eqref{eq:deltaKw} into~\eqref{eq:deltaCw} further yields~\eqref{eq:deltaCw2}, \addtocounter{equation}{1}where ${\bf G}_w = ( \mathbf{I} + \rho \tilde{\mathbf K}_w )^{-1}$. The functional derivative of $C_w$ is then given by
\begin{align}
	&\frac{\delta C_w}{\delta w(p)}\label{eq:grad}\\\notag
	\!\!\!={}&\frac{\rho}{\log 2}\sum_{n=1}^{N}\sum_{n^\prime=1}^{N} {\bf G}_w [n,n^\prime] h\left({\bf r}_{\rm R}^{(n)},\tilde{\bf r}_{{\rm T}}^{(p)}\right) h^*\left({\bf r}_{\rm R}^{(n^\prime)},\tilde{\bf r}_{{\rm T}}^{(p)}\right).
\end{align}
\begin{algorithm}[t]
	\caption{ADF Variational Gradient Ascent}\label{alg:grad}
	\begin{algorithmic}[1]
		\REQUIRE System parameters $\phi_{\rm T}$, $\phi_{\rm R}$, $\theta_{\rm T}$, $\theta_{\rm R}$, and $z_0$, coordinates of antenna elements ${\bf r}_{\rm R}^{(n)}$ and ${\bf r}_{{\rm T}}^{(m)}$, 
		SNR $\rho$, and number of maximum iterations $I$ or stopping threshold~$\varepsilon_{\rm th}$.
		\ENSURE The ADF $w(p)$ that maximizes the rate functional~$C_w$.
		\STATE Initialize the solution as a constant function, i.e., $w^{(0)}(p)=\frac{M-1}{2}$, $\forall p\in [-1,1]$.
		\STATE Build the Gram matrix $\tilde{\mathbf K}_w$ from~\eqref{eq:kernel_p} and~\eqref{eq:arb_chan}.
		\FOR{$i = 0,\cdots,I$}
		\STATE Calculate the gradient by~\eqref{eq:grad}.
		\STATE Update the ADF by~\eqref{eq:updatew}.
		\STATE Apply clipping~\eqref{eq:clipping} and power normalization~\eqref{eq:norm}.
		\IF{$\left\Vert w^{(i+1)}(p)-w^{(i)}(p) \right\Vert_2\leq\varepsilon_{\rm th}$}
		\STATE End iteration.
		\ENDIF
		\ENDFOR
		\RETURN The gradient ascent result $w^{(i+1)}(p)$.
	\end{algorithmic}
\end{algorithm}
With~\eqref{eq:grad} at hand, the original problem can thereby be numerically solved by adopting the gradient ascent method, which iteratively updates the ADF $w(p)$ along the direction of the variational gradient ${\delta C_w}/{\delta w(p)}$ to maximize the rate functional $C_w$. The update rule for $w(p)$ at the $i$-th iteration is given by
\begin{equation}
	w^{(i+1)}(p) = w^{(i)}(p) + \eta \frac{\delta C_w}{\delta w^{(i)}(p)},
	\label{eq:updatew}
\end{equation}
where $\eta > 0$ is the updating step size. Since the ADF must satisfy the constraints in~\eqref{eq:P2}, we apply non-negative clipping
\begin{equation}
	w^{(i+1)}(p) \leftarrow \max\left(w^{(i+1)}(p), 0\right)
	\label{eq:clipping}
\end{equation}
and power normalization 
\begin{equation}
	w^{(i+1)}(p) \leftarrow \frac{(M - 1) w^{(i+1)}(p)}{\displaystyle\int_{-1}^{1} w^{(i+1)}(p) {\rm d}p}
	\label{eq:norm}
\end{equation}
after each iteration\footnote{We can also apply a minimum spacing constraint before power normalization by performing $w(p)\leftarrow \min\left( w(p),M-1 \right)$.}. The overall procedure is summarized in \textbf{Algorithm}~\textbf{\ref{alg:grad}}. 

The computational complexity of this algorithm is dominated by the number of multiplications in~\eqref{eq:grad}, given the specific form of the channel spatial response $h({\bf r}_{\rm R}^{(n)},{\bf r}_{\rm T}^{(m)})$. Suppose the ADF $w(p)$ is discretized and updated over $P = \chi M$ grids, where $\chi \in\mathbb{Z}$, then the complexities of the multiplications and matrix inversion in~\eqref{eq:grad} are ${\mathcal O}(P N^2)$ and ${\mathcal O}(N^3)$, respectively. Given that $P$ scales linearly with $M$ and $N \ll M$, the overall computational complexity of Algorithm~\ref{alg:grad} increases linearly with $M$.

\section{Asymptotic Solutions for Near-Field LoS Channels}
\label{sec:closedform}

The proposed variational gradient ascent method is applicable to arbitrary scenarios with a given channel spatial response function $h(\cdot,\cdot)$. However, the numerical method still involves iterative procedures and provides only limited insights for practical system design. 
In this section, we focus on the near-field channel model, where the paraxial condition~\cite{Miller:00} between UE and BS holds. We reveal the effect of antenna positions on the intrinsic structure of the near-field channel, and derive an asymptotically optimal closed-form solution for the ADF $w(p)$. This solution has the potential to maximize the rate functional in the near-field LoS scenario by revamping a favorable structure of the Gram matrix.
\subsection{Near-Field Channel Model}
In the near-field communication scenario, the spherical wave function, which represents the spatial impulse response of EM waves, can no longer be approximated by plane waves as in the far-field region. Therefore, in this paper we adopt the spherical wave model~\cite{9693928,10845870} to characterize the near-field communication channel.

In near-field channels, the LoS path is less likely to be obstructed and exhibits significantly higher signal strength compared to the non-line-of-sight (NLoS) counterparts~\cite{10845870}. Consequently, the achievable rate of the communication link is predominantly determined by the LoS path. Correspondingly, the near-field channel spatial response is specified by
\begin{align}
	h\left({\bf r}_{\rm T}^{(m)},{\bf r}_{\rm R}^{(n)}\right) = \frac{e^{\jmath \kappa \Vert {\bf r}_{\rm T}^{(m)} - {\bf r}_{\rm R}^{(n)} \Vert }}{\Vert {\bf r}_{\rm T}^{(m)} - {\bf r}_{\rm R}^{(n)} \Vert},
	\label{eq:chmodel_los}
\end{align}
where $\kappa = 2\pi/\lambda$ is the wavenumber with $\lambda$ denoting the wavelength of the carrier.

\subsection{Gram Matrix of Near-Field LoS Channel}

\label{sec:gram}

For notational brevity, we denote $r_{m,n} = \Vert {\bf r}_{\rm T}^{(m)} - {\bf r}_{\rm R}^{(n)} \Vert$ as the distance between the $m$-th antenna element on the BS array and the $n$-th antenna element on the UE array. 
Hence, the Gram matrix of the near-field LoS channel is given by
\begin{equation}
	\tilde{\bf K}_{f}\left[n,n^\prime\right]=\int_{1}^{M}  \frac{e^{\jmath\kappa \left( r_{m,n} - r_{m,n^\prime} \right)}}{r_{m,n}r_{m,n^\prime}}\,{\rm d}{m}.\label{eq:autocorr0}%
\end{equation}
To facilitate further derivations, we introduce different approximations to deal with the distance $r_{m,n}$ in the phase term and denominator of~\eqref{eq:autocorr0}, respectively. For $r_{m,n}$ in the phase term, we introduce
\begin{equation}
	\!r_{m,n} \simeq z_{\rm R}^{(n)} -z_{\rm T}^{(m)}  + \frac{\left( x_{\rm T}^{(m)}-x_{\rm R}^{(n)} \right)^2\!\!+\!\left( y_{\rm T}^{(m)}-y_{\rm R}^{(n)} \right)^2}{2z_0},
	\label{eq:approx_p}
\end{equation}
which is known as the Fresnel approximation~\cite{Miller:00}, and holds as $z_0\gg A_{\rm T}$. While for $r_{m,n}$ in the denominator, we employ another approximation as 
\begin{equation}
	r_{m,n} \simeq z_0-z_{\rm T}^{(m)},%
	\label{eq:approx_d}
\end{equation}
which preserves the elevation angle $\theta_{\rm T}$ in $z_{\rm T}^{(m)}$, in contrast to the coarser yet commonly considered Fraunhofer approximation $r_{m,n}\simeq z_0$~\cite{Miller:00}. 
Substituting~\eqref{eq:approx_p} and~\eqref{eq:approx_d} into the phase and denominator of~\eqref{eq:autocorr0}, respectively, yields
\begin{equation}
	\tilde{\bf K}_{f}\left[n,n^\prime\right] \simeq {\bf D} \overline{\bf K}_{f} {\bf D}^H,
\end{equation}
where ${\bf D}$ is a diagonal matrix containing phase terms as ${\bf D} = {\rm diag}( e^{\jmath \kappa(z_{\rm R}^{(1)}+\frac{( x_{\rm R}^{(1)})^2 + ( y_{\rm R}^{(1)} )^2}{2z_0} )},~\cdots,~e^{\jmath \kappa(z_{\rm R}^{(N)}+\frac{( x_{\rm R}^{(N)})^2 + ( y_{\rm R}^{(N)} )^2}{2z_0} )})$, and $\overline{\bf K}_{f}$ is a real-valued surrogate for its complex-valued counterpart $\overline{\bf K}_{f}$, since it exhibits identical eigen-properties but is defined over $\mathbb{R}$, given by
\begin{align}
	{}&\overline{\bf K}_{f}[n,n^\prime]\notag\\
	={}&\int_{1}^{M} \frac{e^{\jmath\kappa\left(\frac{x_{\rm T}^{(m)} \left( x_{\rm R}^{(n^\prime)}\!-x_{\rm R}^{(n)} \right)+ y_{\rm T}^{(m)} \left( y_{\rm R}^{(n^\prime)}-y_{\rm R}^{(n)} \right) }{z_0}\right) }}{( z_0-z_{\rm T}^{(m)} )^2}{\rm d}m
	\label{eq:autocorr1}
\end{align}
Substituting~\eqref{eq:ant_loc} into~\eqref{eq:autocorr1}, we have 
\begin{align}
	&\overline{\bf K}_{f}\left[n,n^\prime\right]\notag\\
	={}& \int_{1}^{M}\frac{e^{-\jmath\kappa \frac{ A_{\rm T}\Delta n d\sin\theta_{\rm T}\sin\theta_{\rm R}\cos\left( \phi_{\rm T}-\phi_{\rm R} \right)}{z_0} f(m)}}{ ( z_0 - \frac{A_{\rm T}f(m)\cos\theta_{\rm T} }{2})^2} \,{\rm d}{m}\label{eq:sum}\\
	={}& \frac{1}{ z_0^2}\int_{1}^{M}  \frac{e^{-\jmath\beta {\Delta n}  f(m)}}{\left( 1-\tau_{\rm} f(m) \right)^2   }\,{\rm d}{m},\notag
\end{align}
where $\Delta n = n-n^\prime$. {Notably, $\tau = \frac{A_{\rm T}}{2z_0} \cos\theta_{\rm T}$ and 
\begin{equation}
	\beta =  \frac{\kappa A_{\rm T} A_{\rm R} \sin\theta_{\rm T} \sin\theta_{\rm R} \cos\left( \phi_{\rm T}-\phi_{\rm R} \right)}{z_0\left( N-1\right)}
\end{equation}
are two position-related factors, where $\theta_\mathrm{T}$ and $z_0$ represent the elevation angle and distance between the centroids of transceivers, as shown in Fig.~\ref{fig:sysmodel}.} 
To mitigate the non-linearity introduced by $f(m)$ in the phase term of~\eqref{eq:sum}, similar to~\eqref{eq:kernel_p}, we employ integral by substitution as 
\begin{align}
	\overline{\bf K}_{f}\left[n,n^\prime\right] &=\frac{1}{z_0^2}\int_{1}^{M} \frac{e^{-\jmath\beta {\Delta n}  p}}{\left(  1- \tau p \right)^2 }\,{\rm d} f^{-1}(p)\label{eq:nflos_kernel}\\
	&=\frac{1}{z_0^2}\int_{-1}^{1} \frac{w(p)}{\left(  1- \tau p \right)^2} e^{-\jmath\beta {\Delta n} p }\,{\rm d} p\triangleq \overline{\bf K}_w\left[n,n^\prime\right].
	\notag
\end{align}
It can be concluded from~\eqref{eq:nflos_kernel} that the columns of the Toeplitz Gram matrix $\overline{\bf K}_w$ correspond to the \textit{truncated Fourier transform} of the weighted ADF (WADF)
\begin{equation}
	\tilde{w}(p) = \frac{w(p)}{\left(  1- \tau p \right)^2},
\end{equation}
which highlights how the ADF influences the rate functional by shaping the resulting spectral distribution of $\overline{\bf K}_w$.

In fact,~\eqref{eq:nflos_kernel} indicates that the ADF ${w}(p)$ here is closely related to the generating function~\cite{10.1093/imanum/11.3.333}, also known as (a.k.a.) the symbol function~\cite{10.1561/0100000006} of a Toeplitz matrix, which has been widely studied in analyzing asymptotic properties of Toeplitz matrix. In the following subsections, we reveal the connection between the ADF and generating function of a Toeplitz matrices, and derive the optimal ADFs that maximize the asymptotic channel achievable rate functional.

\subsection{Generating Function of the Gram Matrix}
In this subsection, we introduce the concept of generating function and disclose the internal bond between ADF $w(p)$ and the generating function. We first define the generating function in Definition~\ref{def:1}.

\begin{definition}
	\label{def:1}
	Let $g(\theta)$ be a real, smooth, and non-vanishing function defined on unit circle with Fourier coefficients
	\begin{equation}
		c_k = \frac{1}{2\pi} \int_{0}^{2\pi} g(\theta) e^{-\jmath k \theta}~{\rm d}\theta,\ \ k\in\mathbb{Z}.
		\label{eq:fourier_coeff}
	\end{equation}
	If a Toeplitz matrix ${\bf T}_N$ admits the form as
	\begin{equation}
		\renewcommand{\arraystretch}{0.8}
		{\bf T}_N = \left[\begin{matrix}
			c_0&c_{-1}&c_{-2}&\cdots&c_{1-N}\\
			c_1&c_{0}&c_{-1}&\cdots&c_{2-N}\\
			\vdots&\vdots&\vdots&\ddots&\vdots\\
			c_{N-1}&c_{N-2}&c_{N-3}&\cdots&c_{0}
		\end{matrix}\right],
		\label{eq:toeplitz}
	\end{equation}
	then $g(\theta)$ is called the generating function of ${\bf T}_N$.
\end{definition}

Based on~\eqref{eq:nflos_kernel} and Definition~\ref{def:1}, we consider the following Fourier coefficients $c_\ell$ related to the WADF $\tilde{w}(p)$, corresponding to the elements in the Toeplitz matrix $z_0^2\overline{\bf K}_w$, as
\begin{equation}
	\begin{aligned}
		c_\ell = \int_{-1}^{1} \tilde{w}(p) e^{-\jmath \beta \ell p}\,{\rm d}p=\frac{1}{\pi }\int_{-\pi}^{\pi} \tilde{w}\left(\frac{\vartheta}{\pi}\right) e^{-\jmath \beta \ell \frac{\vartheta}{\pi}}\,{\rm d}\vartheta,
	\end{aligned}
	\label{eq:fs}
\end{equation}
where $\ell = \Delta n\in\{ 1-N,2-N,\cdots,N-1 \}$ and $\vartheta = \pi p \in [-\pi,\pi]$. 
Thus, the truncated generating function $g_N(\theta)$ of the Toeplitz matrix $z_0^2\overline{\mathbf{K}}_w$ 
is given by
\begin{align}
	\!g_N(\theta) &= \!\!\sum_{\ell = 1-N}^{N-1} c_\ell e^{\jmath \ell \theta}\notag=\!\frac{1}{\pi}\!\int_{-\pi}^{\pi}\! \tilde{w} \left( \frac{\vartheta}{\pi} \right) \sum_{\ell = 1-N}^{N-1} \!\!\! e^{\jmath \ell \left( \theta- \beta\frac{\vartheta}{\pi} \right)}{\rm d}\vartheta \\
	&= \frac{1}{\pi}\int_{-\pi}^{\pi} \tilde{w}\left( \frac{\vartheta}{\pi} \right)\frac{\sin\left(\left(\frac{2N-1}{2} \right)(\theta-\beta\frac{\vartheta}{\pi})\right) }{\sin\left( \frac{1}{2} (\theta - \beta\frac{\vartheta}{\pi}) \right)}\,{\rm d}\vartheta,\label{eq:symbolf}
\end{align}
which is equivalent to the convolution of a scaled WADF $\tilde{w}(p)$ and the Dirichlet kernel function, a.k.a., the periodic Sinc function. The Dirichlet kernel function becomes the Dirac comb, i.e., the periodic delta function, in the limit
\begin{equation}
	\lim_{N\to \infty} \frac{\sin\left(\left(\frac{2N-1}{2} \right)x\right) }{\sin\left( \frac{1}{2} x \right)} =2\pi \sum_{k\in\mathbb{Z}}\delta\left( x - 2\pi k \right).
	\label{eq:dir}
\end{equation}
Substituting~\eqref{eq:dir} into~\eqref{eq:symbolf}, we have the generating function $g(\theta)$ as
\begingroup
\allowdisplaybreaks
\begin{equation}
	\begin{aligned}
		g(\theta)&=\lim_{N\to\infty} g_N(\theta)\\
		&=2\int_{-\pi}^{\pi} \tilde{w} \left( \frac{\vartheta}{\pi} \right) \sum_{k\in\mathbb{Z}}\delta\left( \theta - \beta\frac{\vartheta}{\pi} - 2\pi k \right)\,{\rm d}\vartheta.
	\end{aligned}
	\label{eq:symbol2}
\end{equation}
\endgroup
Since $\vartheta$ is integrated over $[-\pi,\pi]$, the selective property of Dirac function in~\eqref{eq:symbol2} selects
\begin{equation}
	\theta - \frac{\beta}{\pi}\vartheta -2\pi k = 0,
\end{equation}
which yields 
\begin{equation}
	\vartheta = \frac{\pi}{\beta}\left(\theta-2\pi k \right)\in\left[-\pi,\pi\right] ~~\Leftrightarrow~~-\beta\leq \theta - 2\pi k\leq \beta.
\end{equation}
For typical near-field distance $z_0 > A_{\rm T}$, we have $\beta < \pi$. Recall that $\theta\in[-\pi,\pi)$, which further ensures that a given $k$ contributes to the integral if and only if $k = 0$. 
Hence, all terms with $k\neq 0$ in the summation lie outside the integration interval and thus are discarded. Therefore,~\eqref{eq:symbol2} reduces to
\begin{align}
	g(\theta)
	\!=\!2\!\int_{-\pi}^{\pi}\! \tilde{ w}\left( \frac{\vartheta}{\pi} \right) \frac{\pi}{\beta}\delta\left( \frac{\pi}{\beta}\theta - \vartheta  \right)\,{\rm d}\vartheta
	={}\frac{2\pi}{\beta} \tilde{ w}\left(\frac{\theta}{\beta}\right),%
	\label{eq:symbol3}
\end{align}
where $\theta\in[-\beta,\beta]$. 
Given the generating function $g(\theta)$ of $z_0^2\overline{\mathbf{K}}_w$, the generating function of the Toeplitz matrix ${\bf I} +\rho\overline{\bf K}_w$ in the rate functional~\eqref{eq:P2} is derived as
\begingroup
\allowdisplaybreaks
\begin{align}
		s(\theta) &= \lim_{N\to\infty} \sum_{\ell = 1-N}^{N-1} \left( \mathds{1}\left(\ell,0\right) +\frac{\rho}{z_0^2} c_\ell \right)e^{\jmath \ell \theta}\notag\\
		&=1+\frac{\rho}{z_0^2} g(\theta) =1+\frac{2\pi \rho}{\beta z_0^2}\tilde{ w}\left(\frac{\theta}{\beta}\right),\label{eq:symbol4}
\end{align}
\endgroup
where $\mathds{1}(\cdot,\cdot)$ denotes the indicator function. It is clearly shown that the generating function of ${\bf I} +\rho\overline{\bf K}_w$ is closely related to a \textit{scaled instance of WADF} $\tilde{ w}(p)$.

\subsection{Asymptotically Optimal ADF Design}

In this subsection, we analyze the determinant property of the Toeplitz Gram matrix ${\bf I} + \rho\overline{\bf K}_w$ to derive the optimal ADF $w(p)$, which maximizes the rate functional. {According to~\eqref{eq:nflos_kernel}, the ADF must be well-defined and free of improper singularities\footnote{In this paper, the term singularity collectively refers to zeros (points where the function vanishes) and poles (points where the function diverges).} within $p\in[-1,1]$ to ensure the convergence of the integral~\cite{abramowitz1948handbook}. In practice, some weak or integrable singularities (such as poles with order less than one) may be admissible, which generally leads to well-behaved spectral properties and determinant asymptotics of $\overline{\bf K}_w$\cite{GrenanderSzego1958,fh,pjm/1102968015}.} Therefore, we introduce Fisher-Hartwig conjecture that reveals the relationship between the asymptotic determinant of ${\bf I} + \rho\overline{\bf K}_w$ and its generating function in the following lemma, where the admissible singularities are considered.

\begin{figure}[t]
	\centering
	\includegraphics[width=0.45\textwidth]{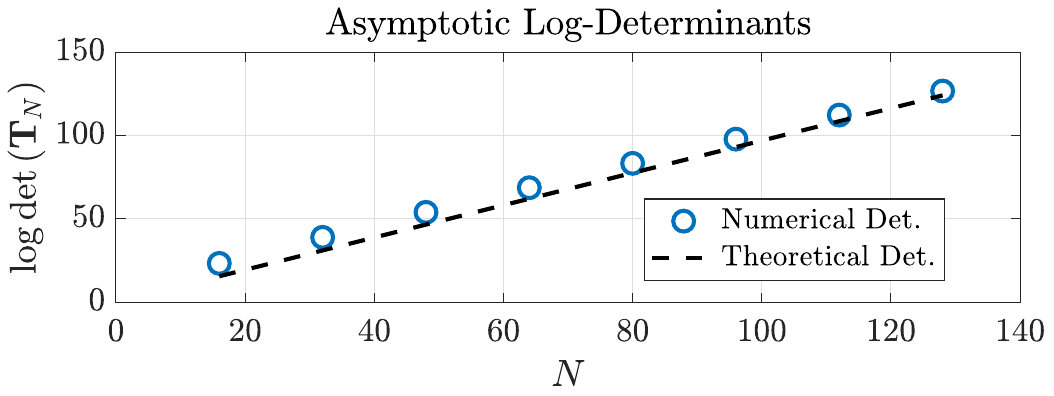}
	
	\caption{The comparison of the exact log-determinant and the asymptotic results with varying values of~$N$.}
	\label{fig:w}
	
\end{figure}
\begin{lemma}[Fisher-Hartwig Conjecture~\cite{fh,pjm/1102968015}]
	Let $s(\theta)$ be a real continuous generating function defined on the unit circle with $R$ singularities that admits the following form
	\begin{equation}
		s(\theta)= b(\theta)\prod_{r=1}^{R} 
		\left|2-2\cos\left(\theta-\theta_r \right) \right|^{\alpha_r},
		\label{eq:fh_symbol}
	\end{equation}
	where $b(\theta)$ is a smooth and strictly positive function on unit circle. For each singularity $\theta_r\in [-\pi,\pi)$, the associated parameters $\{ \alpha_r \}$ satisfy
	\begin{equation}
		\alpha_r > -0.5,\quad\forall r=1,2,\ldots,R,
	\end{equation}
	to ensure convergence. 
	The asymptotic log-determinant of Toeplitz matrix ${\bf T}_N$ associated with generating function $s(\theta)$ is then given by
	\begingroup
	\allowdisplaybreaks
	\begin{align}
		&\lim_{N\to \infty} \log \det \left( {\bf T}_N \right)\notag\\
		={}& E_b\left(\theta\right) + \log N\sum_{r=1}^{R} \alpha_r^2 +\sum_{r=1}^{R} \frac{{\rm BG}^2(1+\alpha_r)}{{\rm BG}(1+2\alpha_r)}\label{eq:fh_det}\\
		&+\sum_{1\leq r<s\leq R} \left\vert2-2\cos\left(\theta_r-\theta_s  \right)  \right\vert^{-\alpha_r\alpha_s} + {\mathcal O}(1)\notag,
	\end{align}
	\endgroup
	where ${\rm BG}(\cdot)$ denotes the Barnes G-function~\cite{10.1145/384101.384104}. The $b(\theta)$-related term $E_b\left(\theta\right)$ is given by
	\begin{equation}
		\begin{aligned}
			&E_b\left(\theta\right)\\ 
			={}&\frac{N}{2\pi} \int_{-\pi}^{\pi}\log b(\theta)\,{\rm d}\theta + \sum_{k=1}^{\infty} k \left( \log b(\theta) \right)_k \left( \log b(\theta) \right)_{-k},
		\end{aligned}
		\label{eq:eb}
	\end{equation}
	where $\{\left( \log b(\theta) \right)_k \}$ denotes the Fourier series of $\log b(\theta)$.
	\label{lemma:fisher_hartwig}
\end{lemma}

\noindent The asymptotic log-determinant~\eqref{eq:fh_det} and the numerical determinant are shown in Fig.~\ref{fig:w}. The tightness of the approximation in~\eqref{eq:fh_det} can be ensured even when $N$ is small, and can be further improved as $N\to \infty$.

{Lemma}~\ref{lemma:fisher_hartwig} reveals that the asymptotic log-determinant in~\eqref{eq:fh_det} is divided into two parts: the $b(\theta)$-related part, i.e., $E_b(\theta)$, contributed by the smooth component $b(\theta)$, and the $\alpha$-related part contributed by the order of singularities. Thus, we categorize the ADF design problem into two cases, i.e., $\alpha=0$ for smooth and non-vanishing ADF, and $\alpha\neq 0$ for ADF with singularities. 

\subsubsection{Case with $\alpha = 0$} In this case, the generating function $s(\theta)$ in~\eqref{eq:fh_symbol} degenerates to the smooth part $b(\theta)$ without singularities. Based on Lemma~\ref{lemma:fisher_hartwig}, the following Corollary can be obtained.
\begin{corollary}
	\label{coro:2}
	Let $b(\theta)$ be a positive, continuous, and smooth function as stated in Lemma~\ref{lemma:fisher_hartwig} that satisfies a total power constraint
	\begin{equation}
		\int_{-\pi}^{\pi} b(\theta)\,{\rm d}\theta = P_0.
		\label{eq:pwrtype}
	\end{equation}
	The term $E_b(\theta)$ in~\eqref{eq:eb} is uniquely maximized when $b(\theta)$ is a constant function.
\end{corollary}
\begin{IEEEproof}
	See Appendix~\ref{append:1}.
\end{IEEEproof}
Corollary~\ref{coro:2} concludes that the optimal generating function $s(\theta)$ of ${\bf I} + \rho\overline{\bf K}_w$ should admit a form of constant function over the interval $[-\beta,\beta]$. Hence, according to~\eqref{eq:symbol4}, the asymptotic achievable rate is maximized when $\tilde{w}(p)$ equals a constant function within $[-1,1]$. Therefore, considering the normalization requirement in~\eqref{eq:pwr_constraint}, the optimal ADF is given by
\begin{equation}
	w(p;\alpha=0) \triangleq \frac{3\left( M-1 \right)}{6+2\tau^2} \left(  1- \tau p \right)^2.
\end{equation}
This result indicates that, for $\alpha = 0$, the optimal ADF depends on the system parameters $M$, $A_{\rm T}$, $z_0$, and $\theta_{\rm T}$.

\subsubsection{Case with $\alpha \neq 0$} 
According to the form of the generating function $s(\theta)$ in~\eqref{eq:symbol4}, the constant bias and non-vanishing $g(\theta)$ guarantee that the $s(\theta)$ does not have a zero-type singularity. 
Therefore, the continuous and differentiable $s(\theta)$ in~\eqref{eq:fh_symbol} can only exhibit pole-type singularities at the edges of the support interval~$[-\beta,\beta]$ with $-0.5<\alpha_r<0$.

In this case, we have $R=2$ pole-type singularities in $s(\theta)$ at the edges of the support $[-\beta,\beta]$, 
i.e., $\theta_1 = -\beta$ and $\theta_2 = \beta$ with $-0.5<\alpha_1,\alpha_2<0$. Then, the asymptotic log-determinant term in the rate functional~\eqref{eq:cap_func} in nats is given by
\begin{align}
		{C}_w ={}&\log \det \left( {\bf I} + \rho \overline{\bf K}_w \right)\notag\\
		={}& E_b(\theta) + \log N\sum_{r=1}^{2} \alpha_r^2 + \sum_{r=1}^{2} \frac{{\rm BG}^2(1+\alpha_r)}{{\rm BG}(1+2\alpha_r)}\label{eq:C_w}\\
		& +\left(2-2\cos2\beta\right)^{-\alpha_1\alpha_2}.\notag
\end{align}
The first-order derivative of~\eqref{eq:C_w} with respect to $\alpha_1$ (or $\alpha_2$ in a similar form) is given by
\begin{align}
	\frac{{\rm d} {C}_w}{{\rm d} \alpha_1} ={}\notag& 2\alpha_1\Bigg(\log N+\frac{{\rm BG}^2(1+\alpha_1)}{{\rm BG}(1+2\alpha_1)}\bigg( 1+{\rm PG}^{(0)}(1+\alpha_1) \\
	&{-2 {\rm PG}^{(0)}(1+2\alpha_1) }\bigg)\Bigg)- \left(2-2\cos 2\beta\right)^{-\alpha_2\alpha_1}\label{eq:firstderiv}\\
	&\times\left(\alpha_2 \log\left( 2-2\cos 2\beta \right)\right),\notag%
\end{align}
where ${\rm PG}^{(0)}(\cdot)$ denotes the Polygamma function of order zero. 
\begin{remark}
	It is important to note that the $\alpha$-related terms in~\eqref{eq:C_w} are strictly positive. This implies that the achievable rate with edge singularities in ADF always exceeds that of its counterpart without such singularities (when $\alpha=0$). In other words, introducing edge singularities leads to a higher achievable rate, as reflected in the achievable rate expression~\eqref{eq:C_w}. Furthermore, as the system dimension tends to infinity as $N\to\infty$, the asymptotic behavior of~\eqref{eq:firstderiv} is dominated by the $\log N$ term, which leads to a negative first-order derivative. Consequently, the asymptotic upper bound of the achievable rate is attained with the specific choice of parameters $\alpha_1 = \alpha_2 \triangleq \alpha = -0.5$.
\end{remark}
\begin{figure}[t]
	\centering
	\includegraphics[width=0.45\textwidth]{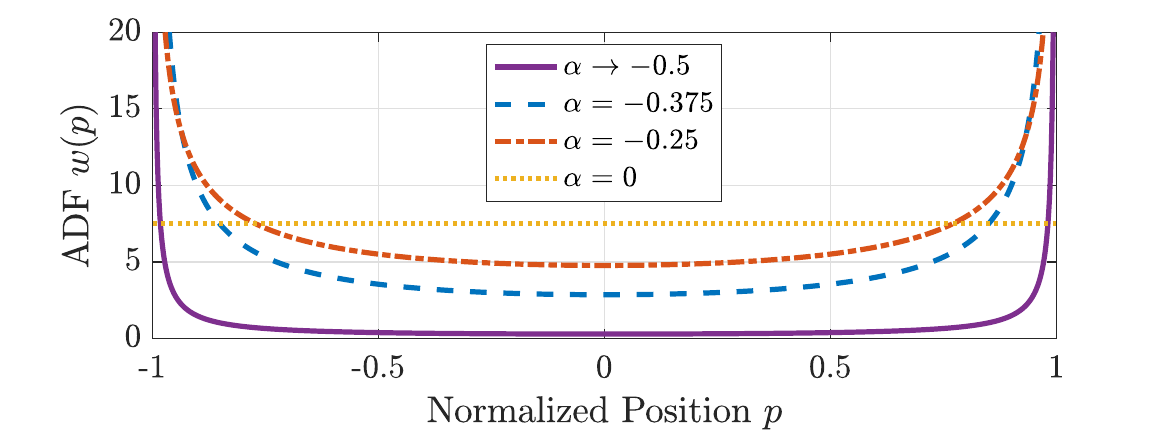}
	
	\caption{Illustrations of ADF $w(p)$ with $z_0 = 10\,$m and $\theta_{\rm T} = \frac{\pi}{2}$.}
	\label{fig:lm2}
	
\end{figure}
Next, we derive the asymptotically optimal ADF with edge singularities. Since $b(\theta)$ must be a constant function according to Corollary~\ref{coro:2}, the generating function in~\eqref{eq:fh_symbol} is then proportional to the product of two trigonometric singularities as
\begin{equation}
	\begin{aligned}
		s(\theta) &\propto \left\vert 1 -\cos\left( {\theta-\beta} \right) \right\vert^{\alpha}\left\vert 1-\cos\left( \theta+\beta \right) \right\vert^{\alpha}\\
		&\propto \left( \cos\beta - \cos\theta \right)^{2\alpha},%
	\end{aligned}
	\label{eq:symbol_func_form}
\end{equation}
which is an even function that is symmetric about $\theta = 0$. Substituting the Maclaurin series $1-\cos\theta = {\theta^2}/{2} + \mathcal{O}(\theta^2)$ into~\eqref{eq:symbol_func_form}, we have
\begin{equation}
	s(\theta) \propto \frac{1}{\left( \beta^2 - \theta^2 \right)^{-2\alpha}}.
	\label{eq:optsymbol}
\end{equation}
Based on the structure of the generating function in~\eqref{eq:symbol4}, the ADF $w(p)$ associated with $s(\theta)$ is given by
\begin{equation}
	\!\!w(p;-0.5\!<\!\alpha\!<\!0) \triangleq \left( \frac{\gamma_\alpha}{\left( 1 - p^2 \right)^{-2\alpha}} \!-\! \frac{\beta z_0^2}{2\pi\rho} \right) (1-\tau p)^2,
	\label{eq:optadf}
\end{equation}
where the normalization constant $\gamma_{\alpha}$ is given by
\begin{equation}
	\gamma_{\alpha}\!=\!\left(M\!-\!1\!+\!\frac{\beta z_0^2 (3+\tau^2)}{3\pi\rho} \right)\! \frac{ \Gamma\!\left(1\!-\!2\alpha\right) \Gamma\!\left( \frac{5}{2} \!+\! 2\alpha \right)\!\sin(2\alpha\pi)}{  \alpha \sqrt{\pi^3}\left(3+4\alpha + \tau^2 \right)}\notag
\end{equation}
to guarantee the constraint in~\eqref{eq:pwr_constraint}, where $\Gamma(\cdot)$ is the Gamma function. 

Different ADFs based on~\eqref{eq:optadf} with distinct values of $\alpha$ are plotted in Fig.~\ref{fig:lm2}. When $\alpha\to0$, the ADF $w(p)$ approaches a constant, corresponding to the conventional ULA configuration. This phenomenon also verifies our theoretical analysis in Corollary~\ref{coro:2}. On the other hand, the optimal ADF $w(p)$ that maximizes the asymptotic achievable rate functional is a $U$-shaped function with poles of order $-0.5<\alpha<0$ at the edges. The more the value of $\alpha$ tends to $-0.5$, the denser antenna deployment is adopted at the edges of the aperture.

Recall that the asymptotic achievable rate upper bound is achieved when $\alpha=-0.5$. In other words, a denser deployment of antenna elements at the edges of the array efficiently captures the higher spatial frequencies of spherical waves at the array periphery, which leads to improved achievable rates. 
While similar insights were observed via numerical simulation in previous works~\cite{7546944,10476966,10643473}, to the best of the authors' knowledge, it is the first time in the literature that it has been disclosed theoretically.

\section{Discrete Antenna Deployment Based on Continuous ADF}
\label{sec:discrete}

In this section, we propose a discretization method to enable the deployment of antennas according to the continuous ADF in closed form. Moreover, to relieve the implementation burden of the potentially high antenna density at the edges, we further propose a flexible deployment strategy for movable antennas.%

\subsection{Direct Discretization}

According to the definition of ADF and its relation with APF in~\eqref{eq:adfdef}, the straightforward way to determine the discrete antenna positions is to find the corresponding APF $f(m)$ and move the $m$-th antenna element to ${\bf r}_{\rm T}^{(m)}$. More specifically, we first find the cumulative ADF (CADF), i.e., the inverse APF, by
\begin{equation}
	\Phi\left( p \right) = \int_{-1}^{p} {w}(p)\,{\rm d}p.
	\label{eq:cadf}
\end{equation}
The position of the $m$-th antenna element can then be obtained by solving the inverse CADF as
\begin{equation}
	p  = f(m) = \Phi^{-1}(m).
	\label{eq:cdfmethod0}
\end{equation}

\begin{figure}[t]
	\centering
	\includegraphics[width=0.45\textwidth]{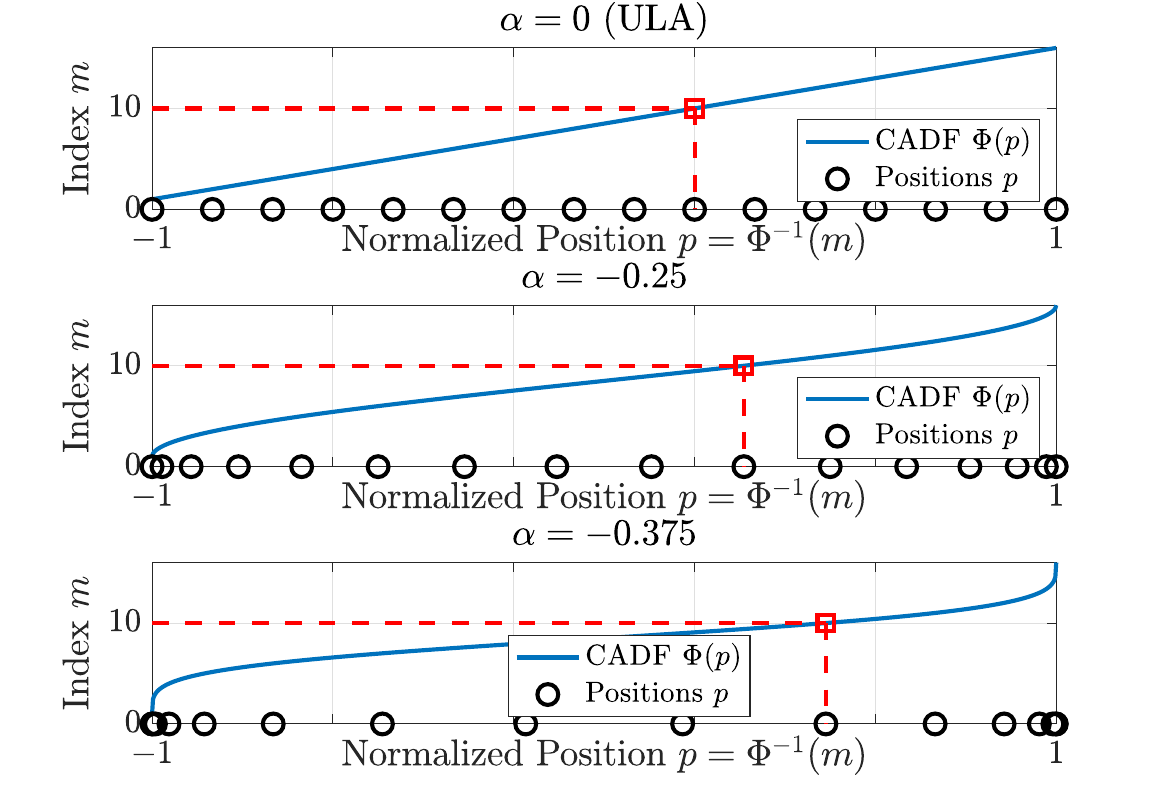}
	
	\caption{Antenna positions on a linear array with normalized aperture $p\in[-1,1]$, $z_0 = 5$ m, $\theta_{\rm T} = \pi/2$ and $M=16$ antenna elements.}
	\label{fig:arr_pos}
	
\end{figure}

It is worth noting that, when massive movable antennas are deployed at the BS and $z_0\gg A_{\rm T}$ is satisfied in the near-field region, the term $(1-p^2)^{-2\alpha}$ dominates the ADF in~\eqref{eq:optadf}. Consequently, ADF further degenerates to
\begin{equation}
	w(p) \simeq \frac{\gamma_\alpha}{\left(1-p^2\right)^{-2\alpha}}.
	\label{eq:adfsimp}
\end{equation}
In this case, the closed-form CADF is available as 
\begin{equation}
	\begin{aligned}
		\Phi\left( p \right) =\frac{M+1}{2}+\frac{\gamma_{\alpha} }{2}B\left(p^2;\frac{1}{2},1+2\alpha\right),
	\end{aligned}
	\label{eq:apf}
\end{equation}
where $B(\cdot;
\cdot,\cdot)$ denotes the incomplete Beta function. Furthermore,~\eqref{eq:apf} yields the closed-form solution of antenna positions as
\begin{equation}
	\begin{aligned}
		p &= \Phi^{-1}(m) = f(m)\\
		&=\sqrt{B^{-1}\left(\frac{2m-\left(M+1\right)}{\gamma_\alpha};\frac{1}{2},1+2\alpha\right)},
	\end{aligned}
	\label{eq:cdfmethod}
\end{equation}
where $B^{-1}\left(\cdot\right)$ denotes the inverse Beta function\footnote{All of the beta functions and their inverses in this paper do not involve interior normalization.}. The antenna positions with different values of $\alpha$ are illustrated in Fig.~\ref{fig:arr_pos}, where CADF is calculated by~\eqref{eq:apf}, and the antenna positions are calculated by~\eqref{eq:cdfmethod}.

\subsection{Flexible Array Implementation}

While in this paper we mainly consider a linear array configuration, it encounters a demerit in practical implementation. 
As illustrated in Fig.~\ref{fig:arr_pos}, the antenna spacing at the edges of the array decreases sharply as $\alpha$ approaches $-0.5$, which may lead to a significant mutual coupling effect and further degrade the system performance by altering the radiation patterns and reducing antenna efficiency. Therefore, direct discretization may not handle the practical constraints properly.

\begin{figure}[t]
	\centering
	\includegraphics[width=0.375\textwidth]{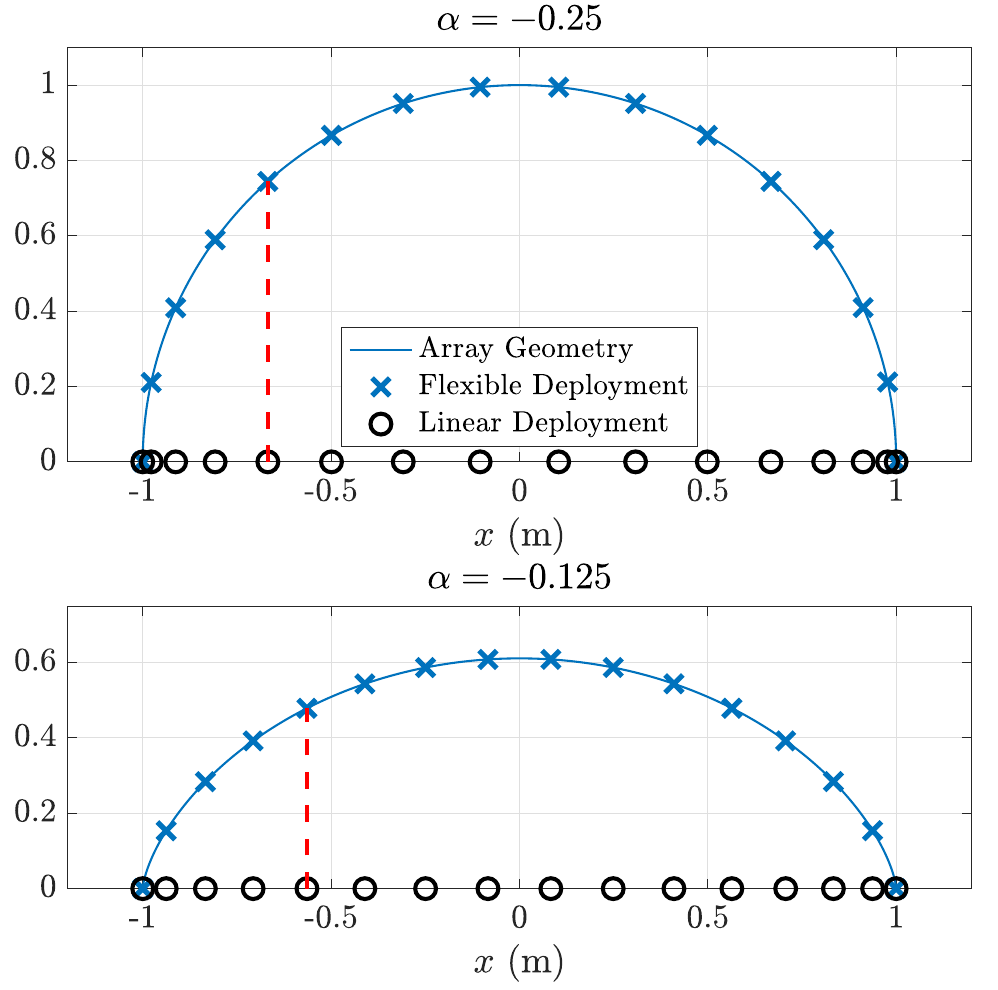}
	
	\caption{Flexible antenna implementations and corresponding projections onto the $x$-axis, with $R=1$ and $M=16$.}
	\label{fig:arr_geo}
	
\end{figure}

One may choose a minimum value of $\alpha$ that guarantees the minimum antenna spacing requirement to strike an effective balance between maximizing achievable rate and reducing mutual coupling. Another appealing alternative is to design a specialized array geometry, which provides more space for antenna deployment and movement to solve the spacing issue. Specifically, we can design a flexible array implementation~\cite{10910066} with antenna elements equally spaced on it, whose curve is described by $(x,y(x))$ and the projection of antenna elements from $(x,y(x))$ onto the $x$-axis aligns with the ADF $w(p)$ for the linear array case. In practical scenarios\footnote{For typical scenarios where the effects of $\theta_{\rm T}$ and $z_0$ are significant, the desired deployment on the flexible array can be determined by applying a central projection~\cite{abramowitz1948handbook} from linear configuration to the flexible array, using $(z_0,\theta_{\rm T})$ as the reference point.} for deriving the ADF in~\eqref{eq:adfsimp}, we can calculate the curve based on it as
\begin{equation}
	w_{\rm flex}(x) \propto \frac{{\rm d}\ell}{{\rm d}x}=\frac{\xi}{\left( R^2-x^2 \right)^{-2\alpha}} = \sqrt{1+\left( \frac{{\rm d}y}{{\rm d}x} \right)^2},
\end{equation}
where $\xi$ is the scaling factor and $R = {A_{\rm T}}/{2}$. At the center of the curve $x=0$, we have ${\rm d}y/{\rm d}x=0$, therefore the scaling factor is $\xi = R^{-4\alpha}$. The curve equation is then given by
\begin{equation}
	\begin{aligned}
		y(x) = \pm\int_{-R}^{x}\sqrt{\left( \frac{R^2-x^2}{R^2} \right)^{4\alpha} -1}\,{\rm d}x + A,%
	\end{aligned}
	\label{eq:curveeq}
\end{equation}
where the bias constant $A$ ensures $y(R) = 0$, i.e., the endpoint of the array is located on the $x$-axis. The flexible array geometries described in~\eqref{eq:curveeq} with different values of $\alpha$ are illustrated in Fig.~\ref{fig:arr_geo}. 

\begin{remark}
	\label{remark2}
	It is worth noting that the achievable rate upper bound-achieving choice of $\alpha = -0.5$ may not be practical, as it will lead to a very large spatial occupation of $(x,y(x))$. 
	For typical scenarios, to ensure feasible constraints such as spatial occupation while maximizing the achievable rate, $\alpha = -0.25$ can be a balanced choice. In particular, this setting yields an isotropic antenna placement, which aligns with a uniform circular array (UCA) configuration~\cite{10900975}. The ADFs are then identical for all directions, which provides a high achievable rate performance and facilitates easier implementation and production.
\end{remark}

\section{Numerical Results}

\subsection{Simulation Setup}

Throughout the simulation, the carrier frequency is set as $f_{\rm c} = 10$~GHz. The UE is equipped with $N=4$ fixed-position antenna elements with the unit spacing being $d=\lambda/2$. The elevation angle of UE is set as $\theta_{\rm R} = \pi/2$, and the azimuth angles are $\phi_{\rm T}=\phi_{\rm R} = 0$. The Rayleigh distance for the BS array is $2A_{\rm T}^2/\lambda = 60\:$m, where $A_{\rm T} = (M-1)d$. 
As the received power is significantly affected by the distance, especially in the near-field region, we normalize the SNR $\rho = 10$ dB. 

\subsection{Impact of Angle and Distance on Antenna Positions}

\begin{figure}[t]
	\centering
	\includegraphics[width=0.45\textwidth]{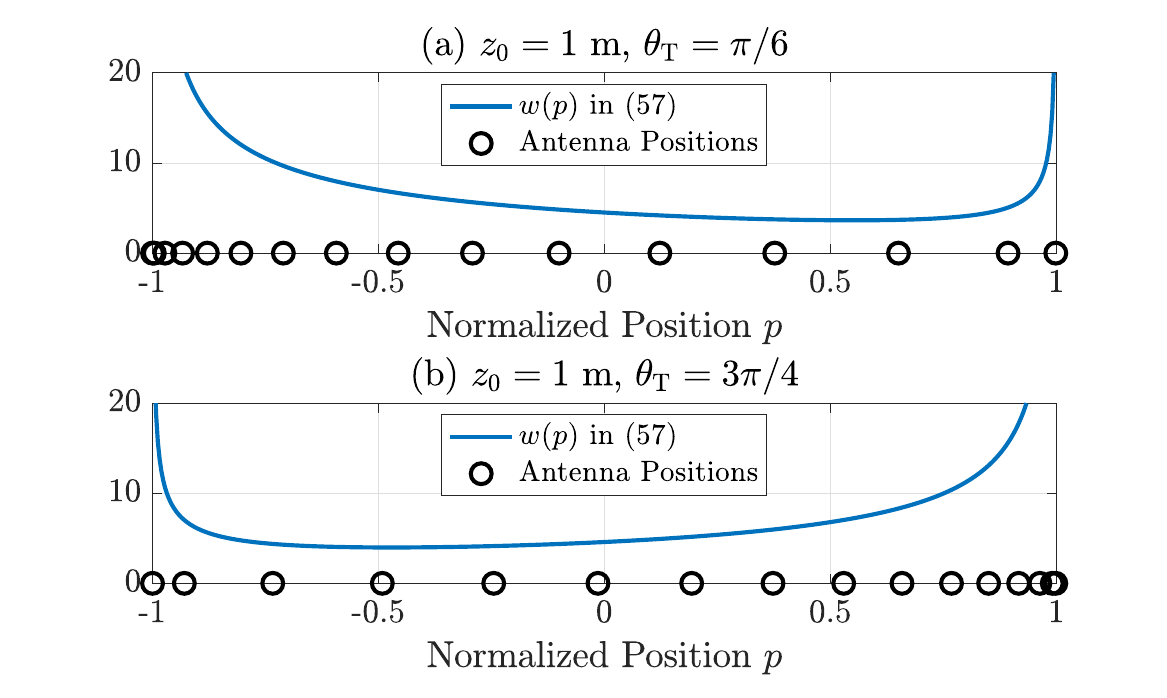}
	\caption{ADF and corresponding discrete antenna positions at $z_0=1$ m.}
	\label{fig:ap_z_theta3}
\end{figure}

\begin{figure}[t]
	\centering
	\includegraphics[width=0.45\textwidth]{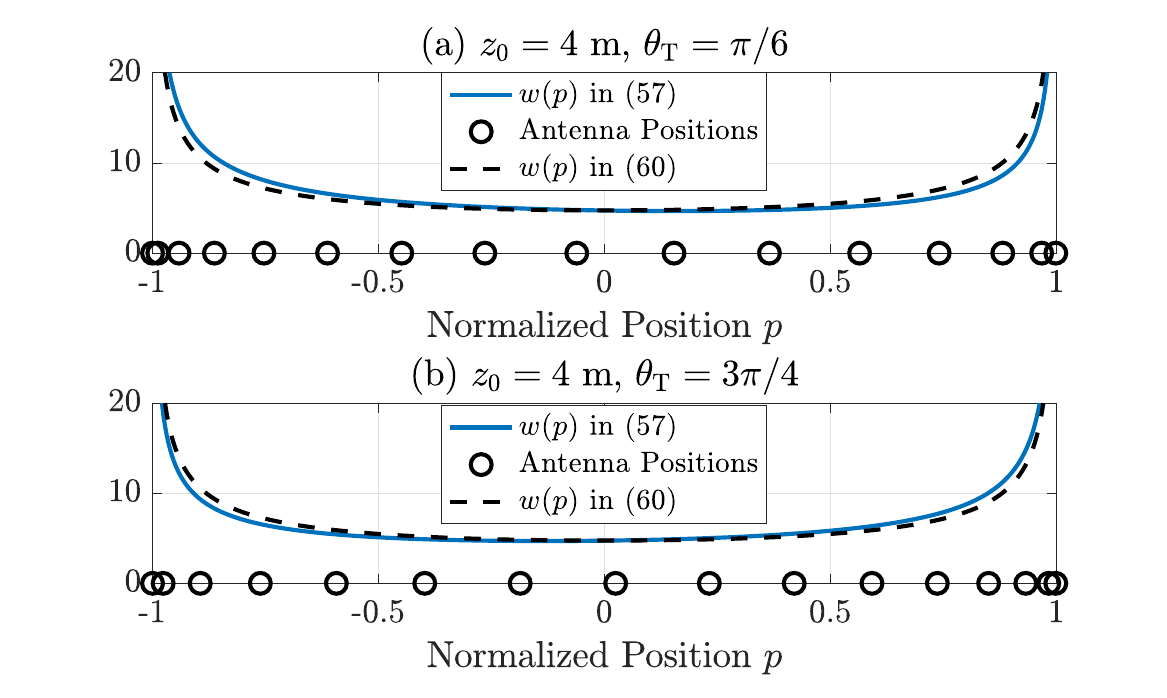}
	\caption{ADF and corresponding discrete antenna positions at $z_0=4$ m.}
	\label{fig:ap_z_theta4}
\end{figure}

In this subsection, we first study the property of the ADF in~\eqref{eq:optadf} for the LoS channel by visualizing the corresponding $M=16$ discrete antenna positions. Demonstrations of the ADFs $w(p)$ and the corresponding antenna positions $p = f(m)$ for $\alpha = -0.25$ are presented in Figs.~\ref{fig:ap_z_theta3} and~\ref{fig:ap_z_theta4}. As shown in Fig.~\ref{fig:ap_z_theta3}, with a centroid distance of $z_0 = 1$ m, the antennas tend to cluster away from the incident directions, highlighting the great importance of \textit{sampling at the far end} in near-field communications. 
This is because the path difference of the EM wave varies more rapidly at the far end of the array, especially in the near-field region, necessitating a higher spatial sampling rate in order to maximize the achievable rate. 

Furthermore, in Fig.~\ref{fig:ap_z_theta4} with $z_0 = 4$ m, the ADFs exhibit reduced skewness compared with Fig.~\ref{fig:ap_z_theta3} with the same $\theta_{\rm T}$ configurations, indicating that the antenna positions become less sensitive to the incident directions. In addition, as can be observed in Fig.~\ref{fig:ap_z_theta4}, two ADFs in~\eqref{eq:optadf} and~\eqref{eq:adfsimp} are almost identical, which implies the effectiveness of the closed-form ADF in~\eqref{eq:adfsimp} and APF in~\eqref{eq:cdfmethod} for antenna position design with large $z_0$ in the near field. More importantly, these results suggest that for larger $z_0$, fixed antenna patterns can closely approach the performance bound, thereby reducing the necessity for direction-dependent antenna position optimization.

\subsection{Proposed Variational Gradient-Based Method}

\begin{figure}[t]
	\centering
	\includegraphics[width=0.45\textwidth]{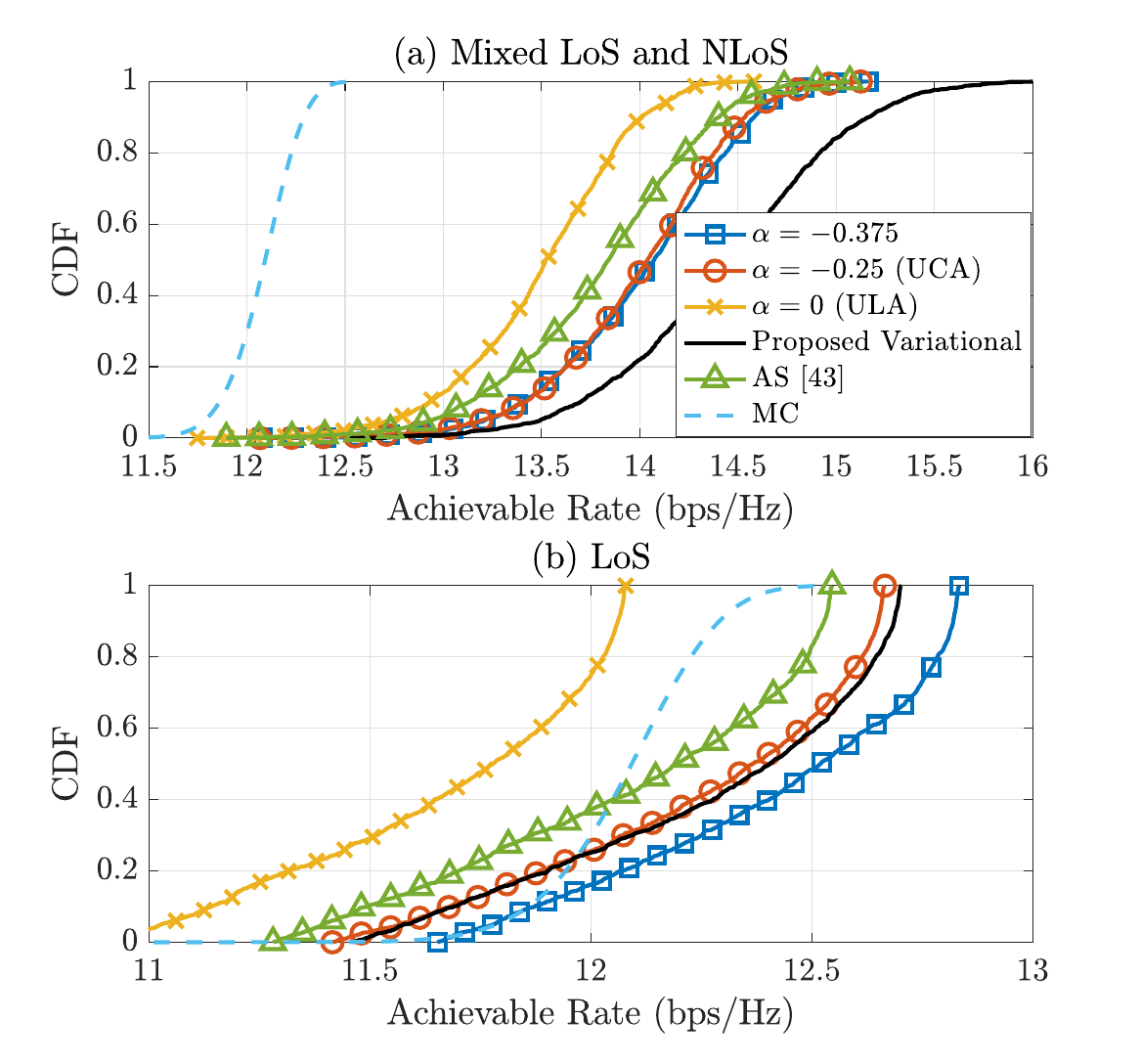}
	\caption{CDF of the achievable rate under the mixed LoS and NLoS near-field channel with $z_0=3$ m and $M=64$.}
	\label{fig:cap_cdf1_2}
\end{figure}

We first investigate the achievable rate achieved by the proposed variational gradient-based method in {Algorithm}~\ref{alg:grad}. 
Recall that this method is applicable to arbitrary channel responses, so we test its performance under a Rician fading channel with mixed near-field LoS and NLoS components as
\begin{equation}
	{\bf H}_{f} = \sqrt{\frac{K}{1+K}}{\bf H}_{{\rm LoS},f} + \sqrt{\frac{1}{1+K}}{\bf H}_{{\rm NLoS},f},
	\label{eq:chmodel_sum}%
\end{equation}
where $K=10$ dB is the Rician factor, while ${\bf H}_{{\rm LoS},f}$ and ${\bf H}_{{\rm NLoS},f}$ denote the LoS and NLoS parts, respectively. The LoS part is modeled by the spherical wave model as~\eqref{eq:chmodel_los}, while the NLoS part comprises $L$ multi-path components (MPCs), each associated with a scatterer located at ${\bf r}_{\rm S}^{(\ell)}$, as

\begin{equation}
	{\bf H}_{{\rm NLoS},f}[n,m] = \sum_{\ell=1}^{L} h\left({\bf r}_{\rm S}^{(\ell)},{\bf r}_{\rm R}^{(n)}\right) h^{*}\left({\bf r}_{\rm T}^{(m)},{\bf r}_{\rm S}^{(\ell)}\right),
\end{equation}
where we have $L = 20$ scatterers uniformly distributed along a circular arc area of radius $3$ m, spanning the angular range over $[\pi/6, 5\pi/6]$. For comparison, we mainly consider the following schemes at the BS side for comparison.
\begin{itemize}
	\item {\bf Proposed Variational}: The optimal ADF is obtained by employing the proposed \textbf{Algorithm~\ref{alg:grad}}, while the positions of antenna elements are discretized by adopting~\eqref{eq:cdfmethod0}. The maximum iteration is set as $I = 50$, and the step size for the gradient update is $\eta = 10^{-3}$.
	\item {\bf Proposed Closed-Form}: The ADF is obtained from the proposed closed-form ADF in~\eqref{eq:adfsimp}, and the positions of the antenna elements can also be calculated in closed-form by~\eqref{eq:cdfmethod}. The range of parameter $\alpha$ varies in $\alpha\in(-0.5,0]$, where $\alpha = 0$ represents the ULA scheme.
	\item {\bf Antenna Selection (AS)}~\cite{10458417}: The antenna positions are selected from ${P}_{\rm AS}=2 M$ uniform grids.%
	\item {\bf Monte-Carlo (MC)}: During each simulation step, $2,000$ random APFs are randomly generated at the BS for achievable rate evaluation.
\end{itemize}
Furthermore, although the closed-form APF in~\eqref{eq:cdfmethod} is derived for the LoS-only scenario, it is intriguing to see its performance in the presence of NLoS components. In this case, we select $\alpha\in\{-0.375,-0.25,0\}$ to construct three closed-form solutions for comparison. %

The cumulative density functions (CDFs) of achievable rates achieved by the proposed variational gradient method and the comparing methods are shown in Fig.~\ref{fig:cap_cdf1_2}(a). As depicted, the 
achievable rate performance of the closed-form solutions increases as $\alpha$ approaches $-0.5$, which verifies our analysis in Section~\ref{sec:gram}. 
However, the proposed variational scheme achieves the highest achievable rate with a comparable standard deviation. This is because {Algorithm~\ref{alg:grad}} is a generic ADF design approach that applies to arbitrary channel responses, and therefore can effectively handle the NLoS paths.

We then evaluate the near-field achievable rate in a LoS scenario. As illustrated in Fig.~\ref{fig:cap_cdf1_2}(b), in the absence of NLoS paths, the overall rate performance is slightly lower than that in the mixed scenarios presented in Fig.~\ref{fig:cap_cdf1_2}(a), and the rate performance of the closed-form solutions improves as $\alpha$ decreases. 
Due to the non-convexity of problem $\mathcal{P}_2$, the proposed variational method can only achieve performance  comparable to the closed-form solution when $\alpha = -0.25$. 
Although its performance does not match the theoretically optimal solution for the LoS scenario, it still provides a significant improvement over the conventional ULA scheme in the near-field region.

\subsection{Proposed Closed-Form Solutions}

\begin{figure}[t]
	\centering
	\includegraphics[width=0.45\textwidth]{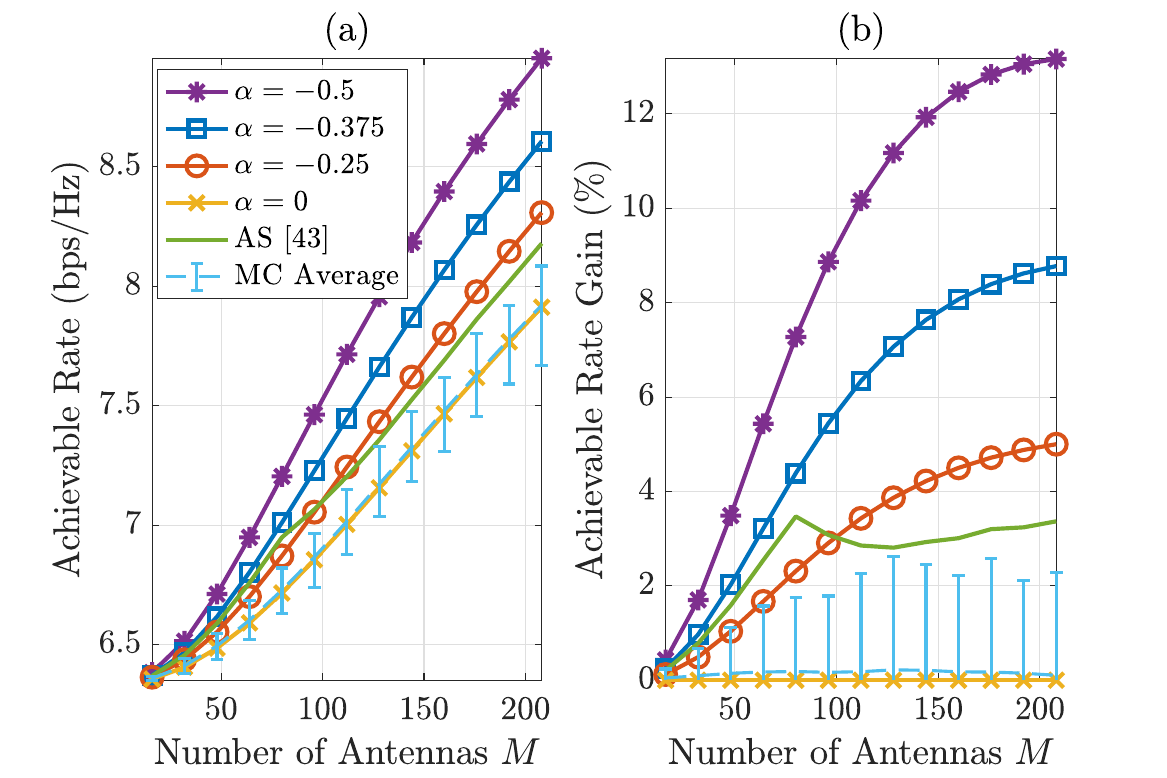}
	\caption{(a) The achievable rate performance with different numbers of antenna elements $M$ on BS array and (b) the corresponding relative performance gain with respect to the ULA fixed antenna scheme.}
	\label{fig:cap_m}
\end{figure}

We then study the achievable rate performance in a LoS channel with a varying number of transmit antennas $M$, when $z_0=5$ m and $\theta_\mathrm{T}=\pi/2$. 
As can be observed in Fig.~\ref{fig:cap_m}, the achievable rate performance increases rapidly with $M$, since a larger number of antennas expand the array aperture $A_{\rm T}$, thereby intensifying near-field effects and enhancing the spatial DoF of the channel. The proposed scheme outperforms both the ULA and AS schemes, exhibiting a faster rate of performance improvement. 
This indicates that the proposed closed-form solution fully exploits the spatial DoFs provided by the additionally deployed movable antennas. As shown in Fig.~\ref{fig:cap_m}(b), the relative achievable rate gain saturates as $M$ further increases, since the number of receive antennas $N$ gradually becomes the main limiting factor of spatial DoF.

\subsection{Complexity Analysis}
The computational complexity of different antenna positioning methods along with the average runtime over $1,000$ Monte-Carlo test cases under the LoS channel model is shown in Table~\ref{tab:complexity}, where the numbers are rounded to three significant figures. Compared with the greedy AS method~\cite{10458417} and alternating optimization (AO) algorithms~\cite{10663427}, the proposed variational method exhibits only linear complexity with respect to the dominant variable $M$, i.e., the number of antennas at the BS side. Therefore, the complexity is comparable to the greedy method when $M$ is small, and grows at a slower rate than the greedy one. The AO algorithm can barely deal with $M=16$ antennas and costs hundreds of seconds to converge, while it is not applicable for $M=128$ massive antennas. Simulation results indicate that Algorithm~\ref{alg:grad} strikes a delicate balance between computational complexity and performance for arbitrary channel conditions, making it a practical choice for large-scale antenna systems. Meanwhile, the closed-form solution demonstrates exceptional computational efficiency, and thus serves as an excellent candidate for extremely efficient movable antenna positioning in the near-field LoS scenario.

\begin{table}[t]
	\centering
	\caption{Computational Complexity (Number of Multiplications) of different Methods.}
	\label{tab:complexity}
	\begin{tabular}{c|c|cc}
		\hline\hline
		\multirow{2}{*}{\bf Methods} & \multirow{2}{*}{\bf \begin{tabular}[c]{@{}c@{}}Computational \\ Complexity  ($\mathcal{O}$)\end{tabular}} & \multicolumn{2}{c}{\bf Ave. Runtime (ms)} \\ \cline{3-4} 
		&                                           &     $\boldsymbol{M=16}$         &      $\boldsymbol{M=128} $          \\ \hline
		{Algorithm~\ref{alg:grad}} &    $I(MN^2+N^3)$                                         &        $58.4$         &      $4.06\!\times\! 10^2$          \\ \hline
		\!\!Closed-form~\eqref{eq:cdfmethod}\!\! &           $\rm N/A$                                &      \!\!\!\!$8.77\!\times\!10^{-3}$\!\!\!\!           &     \!\!\!$5.37\!\times\!10^{-2}$\!\!\!           \\ \hline
		AS~\cite{10458417}&     \!\!\!\!$P_{\rm AS}(N\!+\!MN^2)\!+\!MN^3$\!\!\!\!                                      &   $23.4$              &  $1.68\!\times\! 10^3$              \\ \hline
		AO~\cite{10663427}&     $M^3N^{3.5}$                                      &   $3.52\times 10^5$              &   $\rm N/A$             \\ \hline\hline
	\end{tabular}
\end{table}

\section{Conclusion}
In this paper, we proposed a novel framework to address the antenna position optimization problem with continuous ADF in near-field massive PRA systems. By leveraging functional analysis methods, we proposed a variational method to optimize the ADF in the continuous domain for arbitrary near-field channel models. Furthermore, we derived the optimal form of ADF and proved that the edge density of ADF plays a significant role in maximizing near-field LoS achievable rate. Our results demonstrate that increasing the density of antenna elements at the array edges can effectively enhance the near-field achievable rate. In this sense, a flexible array implementation provided a practical trade-off between spatial constraints and achievable rate maximization, thereby facilitating both higher rate performance and ease of implementation for practical near-field scenarios.

\appendices
\section{Proof of Corollary~\ref{coro:2}}
\label{append:1}
Let $\ell_b(\theta) = \log b(\theta) \geq 0$ for brevity. The Fourier series of $\ell_b(\theta)$ is defined by
\begin{equation}
	l_k = \int_{-\pi}^{\pi} \ell_b(\theta) e^{-\jmath k \theta}\,{\rm d}\theta,
	\label{eq:fourierlb}
\end{equation}
which is a conjugate symmetric sequence since $b(\theta)$ is a real-valued function. Substituting~\eqref{eq:fourierlb} into~\eqref{eq:eb} yields
\begin{equation}
	E_b(\theta) = N l_0 + \sum_{k=1}^{\infty} k\vert l_k \vert^2.
\end{equation}
With the power-type constraint~\eqref{eq:pwrtype}, the Lagrange form for maximizing $E_b(\theta)$ is given by
\begin{equation}
	{L}= N l_0 + \sum_{k=1}^{\infty} k\vert l_k \vert^2 -\mu\left( \vert l_0\vert^2+2\sum_{k=1}^{\infty} \vert l_k \vert^2 - B \right),
\end{equation}
where $\mu$ is the Lagrange multiplier. For $k=0$, the derivative of $L$ with respect to $l_0$ is 
\begin{equation}
	\frac{{\rm d}}{{\rm d} l_0} L = N -2\mu l_0 = 0,
\end{equation}
which yields $l_0 = N/(2\mu)$. For $k\neq 0$, the derivative of $L$ is given by
\begin{equation}
	\frac{{\rm d}}{{\rm d} l_k} L = 2k l_k -\mu 4 l_k,
\end{equation}
which yields $l_k = 0$ or $k=2\mu$. However, since $\mu$ is a globally unique multiplier, the case $k = 2\mu$ does not apply. Therefore, $b(\theta)$ must be a constant function to achieve the maximum of $E_b(\theta)$.
\bibliographystyle{IEEEtran}
\bibliography{IEEEabrv,references}

\end{document}